\documentclass[twocolumn]{aastex63}

\usepackage{graphicx}
\usepackage{epsfig}
\usepackage{epsf}
\usepackage{amsmath}
\usepackage{amssymb}
\usepackage{amsmath, amsthm, amssymb, amsfonts}
\usepackage{xspace}
\usepackage{natbib}
\usepackage{eucal}

\usepackage{colortbl}

\newcommand{\kms}{\ensuremath{\,\rm{km}\,\rm{s}^{-1}}\xspace}

\newcommand{\Msun}{\ensuremath{\,\mathrm{M_\odot}}\xspace}

\newcommand{\Mbh}{\ensuremath{\,M_{\rm BH}}\xspace}
\newcommand{\Mco}{\ensuremath{\,M_{\rm CO}}\xspace}

\newcommand{\chieff}{\ensuremath{\,\chi_{\rm eff}}\xspace}

\newcommand{\Myr}{\ensuremath{\,\mathrm{Myr}}\xspace}
\newcommand{\Gyr}{\ensuremath{\,\mathrm{Gyr}}\xspace}

\newcommand{\COMPAS}{{\tt COMPAS}\xspace}
\newcommand{\python}{{\tt python}\xspace}

\newcommand{\STROOPWAFEL}{{\tt STROOPWAFEL}\xspace}
\newcommand{\Mbhmax}{\ensuremath{\, M_{\rm BH,max}}\xspace}
\newcommand{\Msysmax}{\ensuremath{\,M_{\rm BBH,max}}\xspace}
\newcommand{\Mbhtot}{\ensuremath{\,M_{\rm BBH}}\xspace}
\newcommand{\Mbheen}{\ensuremath{\,M_{\rm BH, 1}}\xspace}
\newcommand{\Mbhtwee}{\ensuremath{\,M_{\rm BH, 2}}\xspace}

\newcommand{\Rmwg}{\ensuremath{\,\mathcal{R}\mathrm{_{MWG}}}\xspace}
\newcommand{\Rvol}{\ensuremath{\,\mathcal{R}\mathrm{_{vol}}}\xspace}
\newcommand{\RvolLIM}{\ensuremath{\,\mathcal{R}\mathrm{_{vol,45}}}\xspace}
\newcommand{\fgaldertig}{\ensuremath{\,f_{\rm30}}\xspace}
\newcommand{\fgalLIM}{\ensuremath{\,f_{\rm45}}\xspace}

\newcommand{\fzestig}{\ensuremath{\,f_{\rm BBH,60}}\xspace}
\newcommand{\fnegentig}{\ensuremath{\,f_{\rm BBH,90}}\xspace}

\newcommand{\revision}[1]{{#1}}

\graphicspath{{./}{figures/}}

\received{\today}

\shorttitle{PISN mass gap events from isolated binaries}
\shortauthors{van Son et al.}

\begin{document}

\title{Polluting the pair-instability~mass~gap for binary black holes \\ through super-Eddington accretion in isolated binaries}


\author[0000-0001-5484-4987]{L.A.C. van Son}
\affiliation{Center for Astrophysics | Harvard \& Smithsonian, 60 Garden Street, Cambridge, MA 02138, USA}
\affiliation{Anton Pannekoek Institute for Astronomy, University of Amsterdam, Science Park 904, 1098XH Amsterdam, The Netherlands}

\author[0000-0001-9336-2825]{S. E. de Mink}
\affiliation{Center for Astrophysics | Harvard \& Smithsonian, 60 Garden Street, Cambridge, MA 02138, USA}
\affiliation{Anton Pannekoek Institute for Astronomy, University of Amsterdam, Science Park 904, 1098XH Amsterdam, The Netherlands}

\author[0000-0002-4421-4962]{F. S. Broekgaarden}
\affiliation{Center for Astrophysics | Harvard \& Smithsonian, 60 Garden Street, Cambridge, MA 02138, USA}

\author[0000-0002-6718-9472]{M. Renzo}
\affiliation{Center for Computational Astrophysics, Flatiron Institute, New York, NY 10010, USA}

\author[0000-0001-7969-1569]{S. Justham}
\affiliation{School of Astronomy and Space Science, University of the Chinese Academy of Sciences, Beijing 100012, China}
\affiliation{Anton Pannekoek Institute for Astronomy and GRAPPA, University of Amsterdam, NL-1090 GE Amsterdam, The Netherlands}

\author[0000-0003-1009-5691]{E. Laplace} 
\affiliation{Anton Pannekoek Institute for Astronomy and GRAPPA, University of Amsterdam, NL-1090 GE Amsterdam, The Netherlands}

\author[0000-0002-8918-5130]{J. Mor\'an-Fraile} 
\affiliation{Heidelberger Institut für Theoretische Studien, Schloss-Wolfsbrunnenweg 35, 69118 Heidelberg, Germany}

\author[0000-0002-8717-6046]{D. D. Hendriks} 
\affiliation{Department of Physics, University of Surrey, Guildford, GU2 7XH Surrey, UK}

 \author[0000-0003-3441-7624]{R. Farmer}
\affiliation{Anton Pannekoek Institute for Astronomy and GRAPPA, University of Amsterdam, NL-1090 GE Amsterdam, The Netherlands}
\affiliation{Center for Astrophysics | Harvard \& Smithsonian, 60 Garden Street, Cambridge, MA 02138, USA}


\begin{abstract}
The theory for single stellar evolution predicts a gap in the mass distribution of black holes (BHs) between approximately 45--130\Msun, the so-called ``pair-instability mass gap". %
We examine whether BHs can pollute the gap after accreting from a stellar companion. To this end, we simulate the evolution of isolated binaries using a population synthesis code, where we allow for super-Eddington accretion. 
Under our most extreme assumptions, we find that at most about 2\% of all merging binary BH systems contains a BH with a mass in the pair-instability mass gap, and we find that less than 0.5\% of the merging systems has a total mass larger than 90\Msun.  
We find no merging binary BH systems with a total mass exceeding 100\Msun. 
We compare our results to predictions from several dynamical pathways to pair-instability mass gap events and discuss the distinguishable features. 
We conclude that the classical isolated binary formation scenario will not significantly contribute to the pollution of the pair-instability mass gap.  The robustness of the predicted mass gap for the isolated binary channel is promising for the prospective of placing constraints on (i) the relative contribution of different formation channels, (ii) the physics of the progenitors including nuclear reaction rates, and (iii), tentatively, the Hubble parameter.

\end{abstract}


\section{Introduction} \label{sec:intro}

Gravitational-wave detections are starting to reveal the properties of the population of merging binary black holes (BBHs). From the first gravitational-wave detections we learned that heavy black holes with masses $\gtrsim 30\Msun$ exist \citep{Abbott+2016_astrophysical_implications, Abbott2018aGWTC, Abbott2018bPop}, which is well above the typical mass for BHs found in X-ray binaries in our galaxy \cite[e.g.][]{Ozel+2010, Farr+2011}. 

These gravitational-wave detections yield unique information about the physics that governed the lives and deaths of their massive stellar progenitors. 
The first ten gravitational-wave detections already support a dearth of BBH mergers with component masses greater than 45\Msun \citep{Fishbach2017, Abbott2018bPop}. It has been suggested that this dearth can be attributed to so-called Pair-instability supernova \citep[or PISN,][]{Belczynski+2016a, Woosley2017, Stevenson+2019}.

The theory of stellar evolution predicts that massive stars can end their lives as BHs with masses up to about $\Mbhmax~\approx45\Msun$ \citep[e.g.][]{Heger+2002,Woosley+2002,Woosley2017, Farmer+2019,renzo2020sensitivity}. 
Progenitor stars \revision{with initial masses between 100\Msun and 140\Msun} which potentially produce BHs with masses above $\Mbhmax \approx 45\Msun$, become unstable due to the production of electron-positron pairs in their cores. 
This leads to the explosive ignition of oxygen, resulting in complete disintegration of the star in a pair-instability supernova \citep[PISN, ][]{Fowler1964, Rakavy1967,Barkat1967, Fraley1968}.
\revision{Photodisintegration prevents the explosion of the most massive progenitors, with final helium cores of $M_{\rm He} \gtrsim 130 \Msun$, thus allowing for BH formation with masses in excess of $130\Msun$ \citep[e.g.][]{Bond1982,Heger+2002}. }
%
Stellar theory thus predicts a gap in the black hole mass function between approximately $45$ and $130\Msun$, referred to as the pair-instability supernova mass gap (or PISN mass gap).

\citet{Farmer+2019} and \cite{renzo2020sensitivity} show that the predictions for the existence and the location of the pair-instability mass gap are remarkably robust. However, uncertainties in the nuclear reaction rates have a significant effect on \Mbhmax, though they merely shift the location of the gap, and do not affect its existence.
The fact that a robust and quantitative prediction exists for the final remnant masses of very massive stars is remarkable given how little is known about the lives of the most massive stars. \\

This prediction of a gap leads to many applications of \Mbhmax.
For example, \cite{Farr2019} \citep[following][]{Schutz1986, Holz2005} argue that if the BH mass distribution is truly shaped by PISN, \Mbhmax could be used as a standard siren for cosmology.
{Farmer et al. (2020, in prep.)} show that the location of \Mbhmax can be used to constrain stellar physics, in particular the uncertain nuclear reaction rate of ${\rm ^{12}C}(\alpha, \gamma)\rm ^{16}O$. 
It has also been suggested that the existence of a mass gap can help to determine the relative contribution of different formation channels to the overall population of BBHs \citep{sedda2020fingerprints,baibhav2020mass}.

\revision{\subsection{The scope of this work} }
In this work, we consider the possibility of forming BBH mergers where at least one of the components has a mass within the PISN mass gap, which we will refer to as ``PISN mass gap events'' hereafter, via the classic isolated binary channel. 
\revision{The classical isolated binary evolution channel for BBH mergers considers the evolution of stars that are born as members of an isolated binary system and experience a common-envelope (CE) phase \citep{Postnov+2014a,Belczynski+2016,Eldridge+2016,Lipunov2017}.}

We compare our results to predictions from dynamical pathways to PISN mass gap events. For this purpose, we adapt an existing population synthesis code, and we allow BHs to accrete mass from a stellar companion assuming the Eddington accretion rate can be exceeded during either a stable mass-transfer phase or during a common envelope event. 
We investigate the implications for the final masses of the merging BBH population. \\

This paper is structured as follows: we give an overview of different pathways to PISN mass gap events in Section \ref{sec: forming PISN gap events}. We describe our simulations in Section \ref{sec:method}. The resulting predictions for populations of BHs are presented in Section \ref{sec:results}. 
We compare our results to predictions from dynamical pathways to PISN mass gap events and discuss the distinguishable features in Section \ref{sec: result compare path}. We then discuss the robustness of our results in Section \ref{sec:discussion}, and provide a summary of our conclusions in Section \ref{sec:conclusions}.


\section{\revision{Forming PISN mass gap events}}
\label{sec: forming PISN gap events}
\revision{The formation of BBH mergers can be broadly divided into two channels, those originating from isolated binary evolution, and those that require dynamical interaction. We provide a brief overview of how each channel may contribute to the pollution of the PISN mass gap.}

\subsection {Forming PISN mass gap events through the classic isolated binary channel}
In this paper we investigate whether the classical isolated binary evolution channel can contribute to the rate of PISN mass gap events.
%
The first born BH in \revision{the classical isolated binary evolution channel} may accrete mass from its companion star as this star evolves and swells to fill its Roche lobe. 

In most population synthesis simulations of compact object mergers, accretion onto the compact object is assumed to be limited by the Eddington rate \citep[e.g.][]{Belczynski+2002,Vigna-Gomez+2018,Neijssel+2019,Spera2019}. \revision{Though exceptions exist, see for example \cite{Belczynski_2008_bhevol} and \cite{Mondal_2019}.}

The Eddington rate is defined as the threshold where radiation pressure from the accretion luminosity halts the inflow of material in the case of spherical accretion. Assuming pure hydrogen accretion;
\begin{equation}
   \dot{M}_{\rm Edd} = \frac{ \rm 4 \pi\, G \,m_p\, M}{\epsilon\, c\, \sigma_{T}} \approx 10^{-8} \left(\frac{M }{\Msun} \right) \quad \left[\frac{\Msun}{\rm yr}\right],
\label{eq: Medd}
\end{equation}
with G the gravitational constant, $m_p$ the proton mass, \rm{M} the mass of the accreting object, c the speed of light, and $\sigma_{T}$ the Thompson scattering cross section \citep{Eddington1926}. We have assumed an accretion efficiency of $\epsilon = 0.1$ \citep{KingRaine2002}.  

When considering BHs that can accrete from a companion star, the duration of the mass transfer phase is typically short (at most of the order of the thermal timescale of the donor star in the case of stable mass transfer and of the order of the dynamical timescale during a common envelope inspiral) and never longer than about 10\Myr, which is longer than the typical lifetime of massive stars. For these short durations the  Eddington limit poses a very severe restriction on the amount of mass that a BH can accrete, as Eq. \ref{eq: Medd} shows. \revision{For example,} a typical BH of $\Mbh \approx 10\Msun$ cannot accrete more than a solar mass in 10\Myr if its accretion is limited at the Eddington rate. \\

Whether or not the Eddington rate poses an absolute limit to the rate at which BHs can accrete is matter of debate. 
First of all, it is based on several idealized assumptions, such as spherical accretion, that are typically not valid. 
If a BH accretes through a `slim' accretion disk, the photons may escape without preventing accretion onto the BH \citep[e.g.][]{Abramowicz+1988, Jiang+2014, Madau+2014a,Volonteri+2015}. At high accretion rates, larger than approximately $\rm 10 \times \dot{M}_{edd}$, photons may be trapped, and advected into the black hole \citep[e.g.][]{Popham1999,Wyithe+2012,sadowski2015, Inayoshi+2016}. 
It is uncertain how accretion proceeds in such cases, but \cite{Inayoshi+2016} argue that mass accretion in excess of $5000$ times the Eddington accretion rate could occur, and can proceed stably. 

Secondly, super-Eddington accretion has been suggested as the most natural explanation  for a wide range of astronomical phenomena.
For example in the context of the rapid growth of super-massive BHs in galactic nuclei \citep[e.g.][ and references therein]{Volonteri+2005,Pezzulli+2016,Johnson+2016}.
But also in the case of ultra luminous X-ray pulsars \citep{Bachetti+2014,Israel+2017} and the galactic source SS 433 \citep[see][ for a review]{Fabrika2004}. 


The uncertainties related to the applicability of the Eddington \revision{rate} pose an uncertainty on the predictions for binary black hole populations and therefore on the robustness of the prediction of the existence of the PISN mass gap. In this work we consider whether and how the possibility of super-Eddington accretion can \revision{lead to PISN mass gap events}.

\subsection{Pathways to pollute the PISN mass gap \revision{requiring dynamical interaction}  \label{sec:intro pathways}}
Various other pathways have been proposed to create PISN mass gap events. Here, we provide a brief overview of these potential alternative pathways. 
%
\subsubsection{Consecutive mergers of BHs} 
The pathway that has been most extensively studied so far with regards to PISN mass gap events involves multiple consecutive mergers of BHs. These may occur in very dense environments where the escape velocities are large, and BBH can form dynamically. High escape velocities are required to retain the BBH-merger product within the formation environment, thereby enabling a consecutive BH merger \citep{Schnittman2007, Baker2007}. \cite{GerosaBerti2019} estimate that an escape speed of about \revision{$\gtrsim$} 50\kms is required to produce PISN mass gap events through consecutive BH mergers.
Promising sites are nuclear star clusters \citep{Antonini2019} and the disks of active galactic nuclei \citep{McKernan2014, McKernan2018, Secunda2019,secunda2020orbital}, where BHs may assemble in migration traps \citep{Bellovary2016, Yang2019, McKernan2019}. 

Globular clusters have also been proposed as sites to create PISN mass gap events through consecutive mergers \citep{Rodriguez2019}. However, their contribution  may be low due to their low escape velocities. 
Globular clusters can only contribute significantly to the production of PISN mass gap events if the BHs are born with low spin \citep{Rodriguez2019}, which minimizes the BBH merger-recoil.

\subsubsection{Fallback of a H-rich envelope}  
An alternative idea involves a star with a final He core mass just below the limit for pulsational pair-instability, i.e. with $\rm M_{He}\lesssim35\Msun$, and an overmassive hydrogen-rich envelope \citep[][]{Woosley+2007,Spera2019}. 
If such a star would (i)~retain a hydrogen envelope that is substantially more massive than 10\Msun until its final stages, and (ii)~the envelope of this star would fall onto the BH, then the total mass of the resulting BH could exceed the PISN limit. 

\citet{Woosley+2007} find BH masses of up to 65\Msun in their models for single stars, when assuming strongly reduced stellar winds and complete fallback of the hydrogen envelope.
\cite{DiCarlo2019b} propose to produce hydrogen-rich progenitors with core masses near the PISN limit through the merger of two stars in a binary system \citep[see also][]{DiCarlo2019a,Vigna-Gomez+2019,Mapelli2019}. They argue that such stellar mergers may be prevalent in globular clusters as the result of dynamical encounters. 

To be detectable as a PISN mass gap event, these BHs need to pair up with another BH, which may be possible inside young stellar clusters \citep{DiCarlo2019a,DiCarlo2019b}. The predictions for this channel are considered to be uncertain because these stellar mergers are not well understood \citep{Justham+2014,Menon+2017} and since it is unclear whether the hydrogen envelope will fall back onto the BH \citep[e.g.][]{Nadezhin1980,Lovegrove+2013,Wu_2018}.

\subsubsection{Accretion from the interstellar medium}
\cite{Roupas2019} explore the limits of BHs fed by the interstellar medium (ISM), based on earlier work from \cite{Leigh2013}. They assume that BHs in young stellar clusters accrete all the gas from their formation environment. These BHs subsequently form BBH pairs in the cluster through dynamical interactions.
Their simulations suggest that it is possible to populate the PISN mass gap through this pathway, although their results depend heavily on the assumed cluster mass and gas density, as well as the gas depletion time-scale.

\subsubsection{Primordial BHs} 
So far we have implicitly assumed the BHs to be of stellar origin. BHs have been hypothesized to be of primordial nature, in which case they are formed as a result of fluctuations in the early Universe \citep{Zeldovich+1966, Hawking1971}. In principle, such BHs could populate the PISN mass gap, since there is no reason to expect a sudden absence or reduction  of BHs in this mass range  \citep[][]{Carr1975,Bird+2016,Sasaki_2016,Raidal2017,Dvorkin2018}. Primordial BHs also have to dynamically find a companion BH to form a PISN mass gap event.

\section{Method} \label{sec:method}
For this study we use the rapid population synthesis code that is part of the \COMPAS suite. A full description of the code can be found in \cite{Stevenson+2017, Vigna-Gomez+2018, Broekgaarden2019}. Here we give a brief summary with an emphasis on the physics relevant for this study. 

\subsection{Initial parameters}
We assume the masses of the initially more massive stellar components (the primary $M_{1}$) are distributed following a \cite{Kroupa2001} initial mass function and draw masses in the range 20 - 150$\Msun$. The binary systems are assumed to follow a uniform distribution of mass ratios ($0.001 \lesssim q = M_{2}/M_{1} < 1.0$) where the lower limit is set by the minimum mass of the initially less massive component (the secondary component, $M_{2} \geq$ 0.1\Msun). The initial binary separations are furthermore assumed to follow a distribution of orbital separations that is flat in the logarithm \citep{Opik1924} in the range $0.01-1000$ AU. Binary systems that fill their Roche lobe at zero age main sequence are discarded. All simulations assume a metallicity of $Z=0.001$, chosen to represent a typical low metallicity environment in which heavy black holes can form \citep{BelczynskiVink2010, Stevenson+2017} and to be consistent with \cite{Farmer+2019}. In Section \ref{ss merger rates}, we discuss why adopting a single metallicity is sufficient for the purposes of this study.    

To optimize computing time, we use the adaptive sampling algorithm \STROOPWAFEL \citep{Broekgaarden2019} to draw the initial parameters of the binaries. This algorithm consists of an exploration phase to draw massive binaries directly from their initial birth distributions.  After this, systems are drawn from reweighted distributions to optimize for the number of systems that end as a BBH that will merge within a Hubble time. In total we evolve $10^{6}$ binaries for each considered model variation. 
This results in approximately $1.4 \times 10^5$ BBH systems in each model.

\subsection{Evolution and mass loss}
We model the evolution of individual binary systems with the algorithms by \citet{Hurley+2000, Hurley+2002} based on evolutionary models by \cite{Pols+1995}. We account for stellar wind mass loss following \citet{Vink+2000,Vink+2001}, \citet{Hamann+1998} and \cite{Vink+2005}, and we assume enhanced mass loss rates in the regime of luminous blue variables following \citet{Belczynski+2010}. 

\subsubsection{Compact objects and supernova kicks}
The remnant mass is modeled as a function of the estimated carbon-oxygen (CO) core mass at the moment of core collapse ($\Mco$). For $\Mco < 30 \Msun$ we use the delayed model from \cite{Fryer+2012} to determine the remnant masses. For $\Mco > 30 \Msun$ we use the remnant mass prescription from \cite{Farmer+2019} to account for the effects of pair pulsations and pair-instability supernovae (see Appendix \ref{app: farm fry} for a comparison of these two prescriptions). 
With this implementation the lower edge of the pair-instability mass gap is located at $\Mbhmax \approx 43.5$\Msun, for a metallicity $Z=0.001$. 

To model supernova kicks, we draw kick velocities with random isotropic orientations and kick magnitudes from a Maxwellian distribution \citep{Hobbs+2005}. BH kicks are subsequently reduced. For BHs resulting from progenitors with $\Mco < 30 \Msun$ at the moment of core-collapse, BH kicks are reduced by the amount of mass falling back onto them during the explosion mechanism, following \cite{Fryer+2012}. \revision{Since the most massive BHs are thought to form without a supernova explosion, we assume no supernova kick occurs for BHs resulting from progenitors with $\Mco > 30 \Msun$.}

\subsubsection {Mass transfer}
We account for mass transfer when a star overflows its Roche lobe, where the Roche-lobe radius is approximated following \citet{Eggleton1983}. To determine whether Roche-lobe overflow is stable we use an estimate for the response of the radius of the donor star and its Roche lobe as a result of mass transfer \citep[see e.g.][and references therein]{Vigna-Gomez+2018}. 

During stable mass transfer onto a stellar companion we assume that the accretion rate is limited to at most ten times the thermal rate of the accreting star \citep{Neo+1977,Hurley+2002}. Material lost from the system is assumed to carry the specific orbital angular momentum of the accreting star \citep[e.g.][]{Soberman+1997,van-den-Heuvel+2017}. 

Unstable mass transfer is assumed to result in CE evolution \citep{Paczynski1970,Ivanova+2013}. Successful CE ejection is allowed for donor stars that are in the Hertzsprung gap \citep[the optimistic approach to CE, following][]{belczynski2017evolutionary}.\revision{This is consistent with \citet{Stevenson+2019}.}
We assume this shrinks the orbit following the $\alpha, \lambda$ formalism as proposed by \cite{Webbink1984} and \cite{de-Kool1990}, using the fits provided by \citet{Xu+2010a,Xu+2010} that account for the internal energy of the envelope.
If the donor star overflows its Roche lobe directly following a CE event, we assume the binary was not able to eject its envelope and presume the system ends as a stellar merger.

\subsection{Treatment of black hole accretion in this study}
\label{ss: acc onto BHs}
Here we consider different modes where we allow for the possibility of super-Eddington accretion onto BHs, as we describe below. 
%
We adopt the assumption of Eddington limited accretion in what we will refer to as our fiducial simulation (model~0).  Specifically, we limit the accretion onto compact objects to the Eddington rate as given in Eq. \ref{eq: Medd}. 

In our first model variation (model~1) we allow for super-Eddington accretion during phases of stable mass transfer when the accretor is a BH. We consider the extreme limit where the black hole accretes all the mass provided by the donor star.  

In our second model variation (model~2), we consider the accretion of mass onto BHs during the inspiral phase of a CE event.
Following the arguments first presented in \cite{Chevalier1993,Brown1995, BetheBrown1998} and later \cite{MacLeod2015}, the mass accreted by the BH, $\Delta M_{\rm acc}$,  can be estimated as Hoyle-Littleton accretion rate $ \dot{M}_{\rm HL}$ times the duration of the inspiral time, $\Delta t_{\rm insp}$. This gives  

\begin{align}
\begin{split}
\label{eq: BHL accretion rate}
    \Delta M_{\rm acc}  \approx \dot{M}_{\rm HL} \, \Delta t_{ \rm insp}  \approx  \frac{M_{\rm BH, birth} \cdot M_{ \rm comp}}{2\,(M_{\rm BH, birth} + M_{ \rm comp})}, 
\end{split}
\end{align}
where $M_{\rm BH, birth}$ is the birth mass of the BH and  $M_{\rm comp}$ is the mass  of the companion. 
Equation~\ref{eq: BHL accretion rate} approximates the inspiral time as the ratio of the orbital energy to the drag luminosity \citep[i.e.][]{Iben1993}. Unlike \cite{MacLeod2015}, we do not restrict the accreted mass due to microphysics or the envelope structure, implying that our estimates for the final BH masses can be taken as extreme upper limits. 

Lastly, we also run a combined model (model~3) which allows for super-Eddington accretion onto BHs during both stable mass-transfer phases and CE phases. 

The assumptions adopted in our model variations are extreme by design. This allows us to place an upper limit on the masses that stellar-mass BHs can reach. These assumptions are intended to provide upper limits to the contribution of the isolated binary evolutionary channel to PISN gap merger events.

We refer to our first model variation as model 1: `stable accretion model', and to our second model variation as model 2: `CE accretion model'. The combined model variation is referred to as model 3: `combined model'.

Throughout this paper we will use  `PISN mass gap systems' as a shorthand for BBH systems with at least one component with $\Mbh > \Mbhmax$. If a PISN gap system will merge within a Hubble time due to gravitational waves, we refer to it as a `PISN mass gap event'.

We adopt $\Mbhmax = 45\Msun$ for the lower edge of the PISN mass gap. \revision{This value is slightly higher than the \Mbhmax resulting from the simulations by \cite{Farmer+2019} at Z = 0.001, whose prescriptions we adopt to model the final remnant masses. $\Mbhmax = 45\Msun$ is thus chosen to represent a conservative limit for the lower edge of the PISN mass gap. This value is also consistent with the limit used by \cite{Fishbach2019}.}

\section{Results} 
We describe our results for the individual component masses of BBH systems in Section~\ref{ss comp mass},\revision{ and the effect on the mass ratios in Section~\ref{ss results mass ratio}. The distribution of total BBH masses is discussed} in Section~\ref{ss total mass dist}, and the estimated merger rates in Section~\ref{ss merger rates}.

\label{sec:results}
\subsection{Component masses} 
\label{ss comp mass}

\begin{figure*}[ht]
   \centering
    \includegraphics[trim=5cm 9.5cm 5cm 7cm, clip, width=0.9\textwidth]{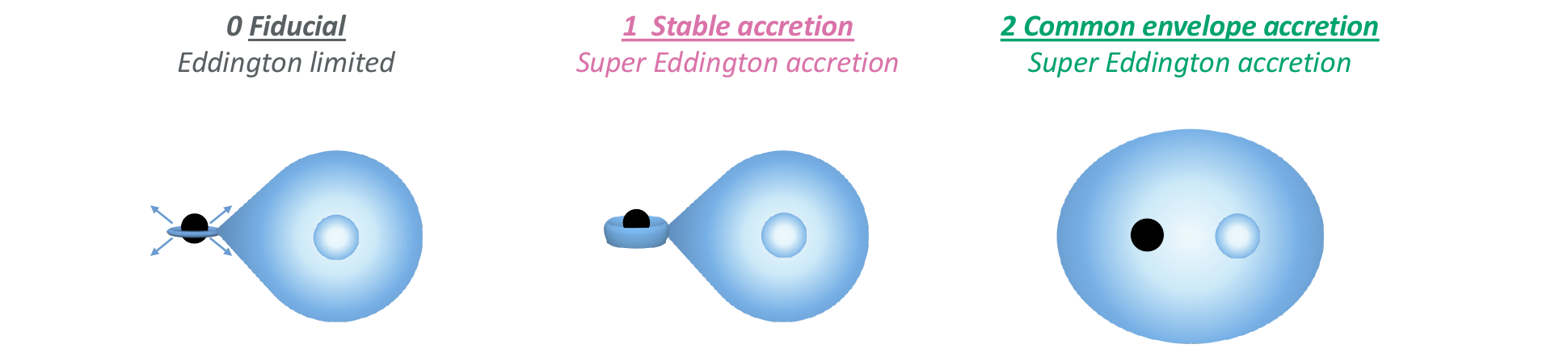} 
    \includegraphics[width=\textwidth]{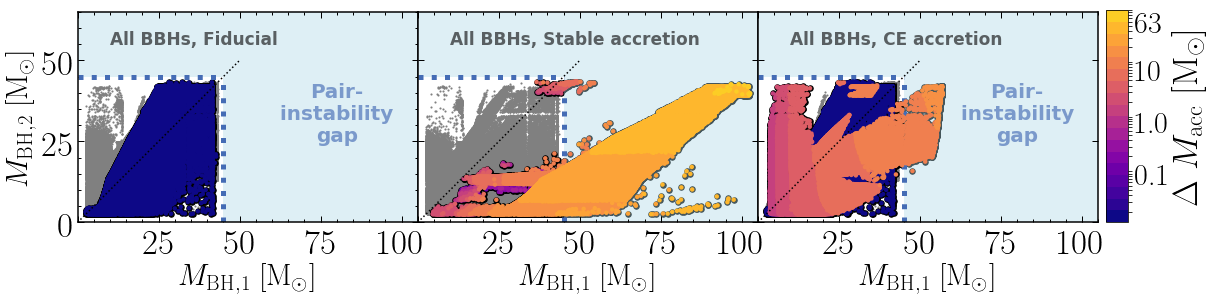} 
    \includegraphics[width=\textwidth]{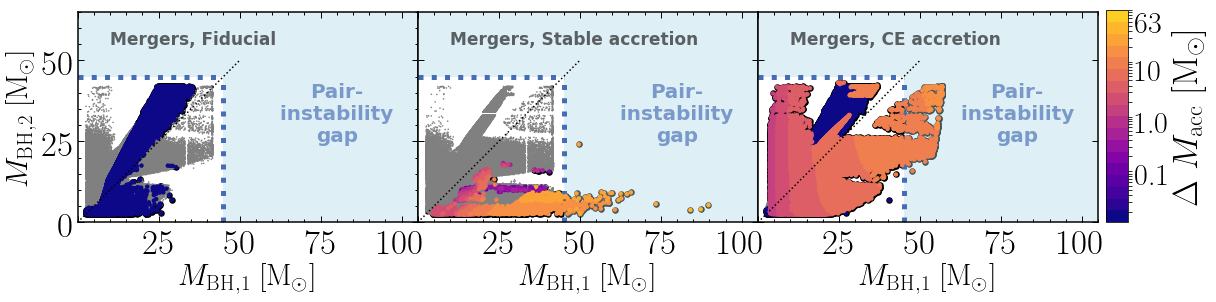} 
    \caption{\textbf{Top:} Cartoon depictions of the BH accretion phase of binary evolution that is varied between the simulations. \textbf{Middle:} The final component masses of the simulated BBH systems for the fiducial population (left column), the BH population that accretes at super-Eddington rates during stable mass transfer (middle column), and the BH population that accretes during CE events (right column). \revision{The light blue shaded region bordered by dotted blue lines indicates the approximate location of the PISN mass gap.} Colors indicate the amount of mass accreted by one of the BH components. Gray dots are systems where the BHs did not accrete any mass. The black dotted line shows where \Mbheen = \Mbhtwee. \textbf{Bottom:} same as middle, but only including the BHs that merge within a Hubble time. Note that the BHs as shown are all those that occur in our simulations; they are not weighted by their formation probability.}
    \label{fig:m1m2}
\end{figure*}

\begin{table*}[t]
\caption{ Comparison of the full population of BBH (All) and the subset of those that merge within a Hubble time (Merging). We provide the maximum BH (\Mbhmax) for individual BHs and the fractions of BBHs with at least one component more massive than 30 and 45\Msun (\fgaldertig and  \fgalLIM, respectively). The errors shown are an estimate of the 1--$\sigma$ errors that result from the statistical sampling uncertainty\revision{, see Appendix \ref{app: merger rate} for the derivation.} }
\label{tab: masses}
\begin{flushleft}
\begin{tabular}{lccc|ccc}   
   &              \multicolumn {3}{c|}{All}     & \multicolumn{3}{c}{Merging}   \\
                                &\fgaldertig                    &\fgalLIM                       &\Mbhmax    &\fgaldertig            &\fgalLIM           & \Mbhmax      \\  
Model                           &  \%                           & \%                            & ($\Msun$) &  \%                   &  \%               &($\Msun$)      \\ \hline \hline  
  0. Fiducial                   & $35.5\pm0.4$                  & $0.0$                         & 43        & $10.7\pm0.1$          & $0.0$             & 42           \\
  1. Stable super-Eddington     & $44.0\pm\revision{0.5}$       & $20.3\pm\revision{0.2}$      & 103       & $7.89\pm0.1$          & $0.13\pm0.03$      & 90           \\%
  2. Common envelope accretion  & $36.6\pm0.4$                  & $0.63\pm0.01$                 & 57        & $14.8\pm0.1$          & $1.33\pm0.05$      & 57           \\
  3. Combined                   & $46.1\pm0.4$                   & $20.7\pm\revision{0.2}$     & 103       & $16.8\pm0.1$          & $2.35\pm0.06$      & 90           \\
\end{tabular}
\end{flushleft}
\end{table*}

Figure \ref{fig:m1m2} shows the distribution of individual BH masses for our fiducial (model variation 0), stable accretion (model variation 1) and CE accretion model (model variation 2). The top row of Figure \ref{fig:m1m2} shows a cartoon depiction of the model variations considered here.
The primary BH mass, \Mbheen, refers to the mass of the BH that originates from the initially more massive star in the binary system. Similarly the secondary BH mass, \Mbhtwee, refers to the mass of the BH that originates from the initially less massive star. 
The middle row displays all BBHs resulting from our simulations, while the bottom row focuses on BBH systems that merge within a Hubble time due to gravitational-wave emission.

BBH systems are shown as gray points unless the first born BH accreted from its companion. For the latter systems, the colors indicate the amount of mass that is accreted by the first born BH through accretion from its stellar companion, $\Delta M_{\rm acc}$. 

We furthermore estimate the fraction of BBHs with at least one component more massive than 30 and 45\Msun, denoted as $\fgaldertig$ and  $\fgalLIM$ respectively. We also quote the maximum mass for individual BHs (\Mbhmax) created in our simulations.  
The results for all three model variations as discussed below are summarized in Table~\ref{tab: masses}.

\subsubsection{Fiducial model} 
The fiducial population does not produce any BBH systems with component masses above $\Mbhmax = 45\Msun$ (i.e. $\fgalLIM= 0\%$), in agreement with earlier studies \citep[e.g.][]{Belczynski+2016a, Stevenson+2019}.  This can be seen in the left-most column of Figure \ref{fig:m1m2} and in Table~\ref{tab: masses}. In practice we find no BHs more massive than $\Mbh \approx 43\Msun$, which is the limit set by the remnant mass function as adopted in this work (see Appendix \ref{app: farm fry}). 

We see that BBHs that have accreted mass (blue points in the left column of Figure \ref{fig:m1m2}) span the whole mass range, but have a slight preference for equal mass ratios.
BHs that are part of a merging BBH system prefer lower primary BH masses, \Mbheen. The BBH systems with the lowest mass \Mbheen primarily result from systems that interacted early on in their stellar evolution. Further substructure in the \Mbheen, \Mbhtwee distribution is caused by their origin from different evolutionary channels \citep[see e.g.][for a discussion of different evolutionary channels]{Dominik+2012}. 


The fraction systems containing a heavy BH, $\fgaldertig$, is about a third for the full population and one tenth for the population that merges. These values are relatively large, this results from the fact that we have assumed a low value for the metallicity $Z= 0.001$, which leads to reduced mass loss through stellar winds and the formation of heavier black holes. This has been pointed out in earlier studies \citep[e.g.][]{Belczynski+2010, Stevenson+2017}.  

The Eddington limit severely restricts the amount of mass that is accreted, $\Delta M_{\rm acc}$, and BHs accrete less than about $0.01 \Msun$ in this model. 
This confirms that accretion cannot lead to PISN mass gap systems in the fiducial model.

The results obtained with our fiducial model are very similar to those presented in \cite{Stevenson+2019}, who also used the \COMPAS suite, with very similar assumptions and initial conditions. Small differences arise from the different treatment of pulsational mass loss and PISNe, which are discussed in Section \ref{sec:method} and Appendix \ref{app: farm fry}).

\subsubsection{Model variation 1: Stable Accretion}
Our first model variation, where we allow for super-Eddington accretion rates during stable mass transfer, is shown in the central column of Figure \ref{fig:m1m2}.
The population of BHs experiencing stable mass accretion accretes between $0.1$ and approximately $63\Msun$, placing many BHs in the PISN mass gap. 
Note that the progenitors of accreting BHs are commonly the initially more massive stars, since these typically evolve on a shorter timescale. We find that approximately one fifth of all systems have at least one BH more massive than 45\Msun, i.e. $\fgalLIM \approx 20\%$. 

The maximum amount of mass that BHs can accrete is limited by the available matter rather than the accretion rate in this model. In practice the available matter equals the mass of the donor's envelope. 
The theoretical maximum for the most massive BH in this model variation is therefore the maximum mass of a BH at birth, plus the maximum envelope mass of the companion. The most massive BH formed in this simulation is just over 100\Msun (i.e., $\Mbhmax \approx 103\Msun$), but we note that for the most extreme masses, our simulations are affected by uncertainties resulting from sampling effects.  

The distribution shows a clear upward diagonal trend, similar to the fiducial simulation but shifted to higher masses for \Mbheen.  This can be understood when considering that higher-mass BHs generally come from higher-mass progenitors, which typically have higher-mass companions. Higher-mass companions typically have more massive envelopes and thus have more mass available to donate to the first born BH, leading to a larger amount of accreted mass, $\Delta M_{\rm acc}$, and a higher-mass primary BH, \Mbheen. At the same time, the higher-mass companions have larger cores and result in higher-mass secondary BHs, \Mbhtwee. 
Thus, the accreted mass ($\Delta M_{\rm acc}$) scales with the final mass of the secondary BH, $\Mbhtwee$. 

Outliers to this main trend exist, as can be seen in the central column, middle row of Figure \ref{fig:m1m2} around $\Mbheen \approx \Mbhtwee \approx 50 \Msun$, and around  $\Mbheen \approx 75 \Msun$ with $ \Mbhtwee \leq 25 \Msun$.
In these cases, the massive stellar progenitor of the BH has already lost most of its envelope due to winds. This occurs in systems where the mass transfer happens at a later evolutionary stage of the donor star \citep[i.e., case C mass transfer,][]{Lauterborn1970}. \\

For conservative mass transfer, as we assume for accreting BHs in this model variation, mass and angular momentum conservation dictate that the binary orbit widens when the donor is less massive than the accretor, i.e. when the mass ratio is reversed \citep[e.g.][]{Soberman+1997}. This is also true when we consider lower accretion efficiencies and assume that the mass that is not accreted is lost with the specific orbital angular momentum of the BH (see Appendix \ref{app: orbital evolution} and Section \ref{ss: var in MT} for a discussion). We find that the mass ratio is almost always reversed by the super-Eddington accretion in this model variation, and thus also that the binary orbit widens in almost all cases. 

Stable mass transfer thus widens BBH systems. Moreover, the more mass is transferred, the wider the system becomes and therefore stable super-Eddington accretion widens BBH systems significantly. Widening the orbit has a strong effect on the gravitational-wave merger time, since it scales with the binary separation to the fourth power \citep{Peters1964}. Sufficiently wide BBH systems cannot merge \revision{within a Hubble time} through gravitational waves alone. \\

The most massive BH that is part of a BBH system that still merges within a Hubble time has a mass of $\Mbh \approx 90\Msun$. The most massive systems ($\Mbheen \geq 70\Msun$) in the stable accretion model that still merge within a Hubble time experience a similar evolution. Merging BBH systems after a long phase of stable super Eddington accretion is only possible through a BH-kick from the secondary BH. A `lucky' BH-kick can increase the binary eccentricity to nearly 1, which radically reduces the gravitational wave inspiral time \citep{Peters1964}. Significant BH-kicks are only implemented in our simulations for relatively low mass CO cores ($\leq 15\Msun$).   Therefore this evolutionary pathway is only possible for relatively extreme mass ratio BBHs. Whether such BBH systems exist in nature furthermore depends on the physics of BH-kicks, which is a matter of debate.


The vast majority of the affected systems do not merge within a Hubble time (central column, bottom row of Figure \ref{fig:m1m2}).
Only 0.1\% of the merging BBH systems in this model variation contain a BH with $\Mbh \geq 45 \Msun$ ($\fgalLIM = 0.1\pm0.03\%$). Moreover, the bulk of PISN mass gap systems created in this model variation is not only too wide to merge within a Hubble time, but is also too wide to be detectable through all planned gravitational-wave detectors such as LISA \citep[based on values from ][]{WeiTou2018}.


\subsubsection{Model variation 2: common envelope accretion} 
Our second model variation, which allows super-Eddington accretion onto BHs during the inspiral phase of a CE event, is displayed in the right-most column of Figure~\ref{fig:m1m2}. This model does produce BHs with masses in the PISN gap, but the BHs are not as massive as those resulting from our stable super-Eddington accretion model.

\begin{figure}[t]
    \centering
    \includegraphics[width=0.45\textwidth]{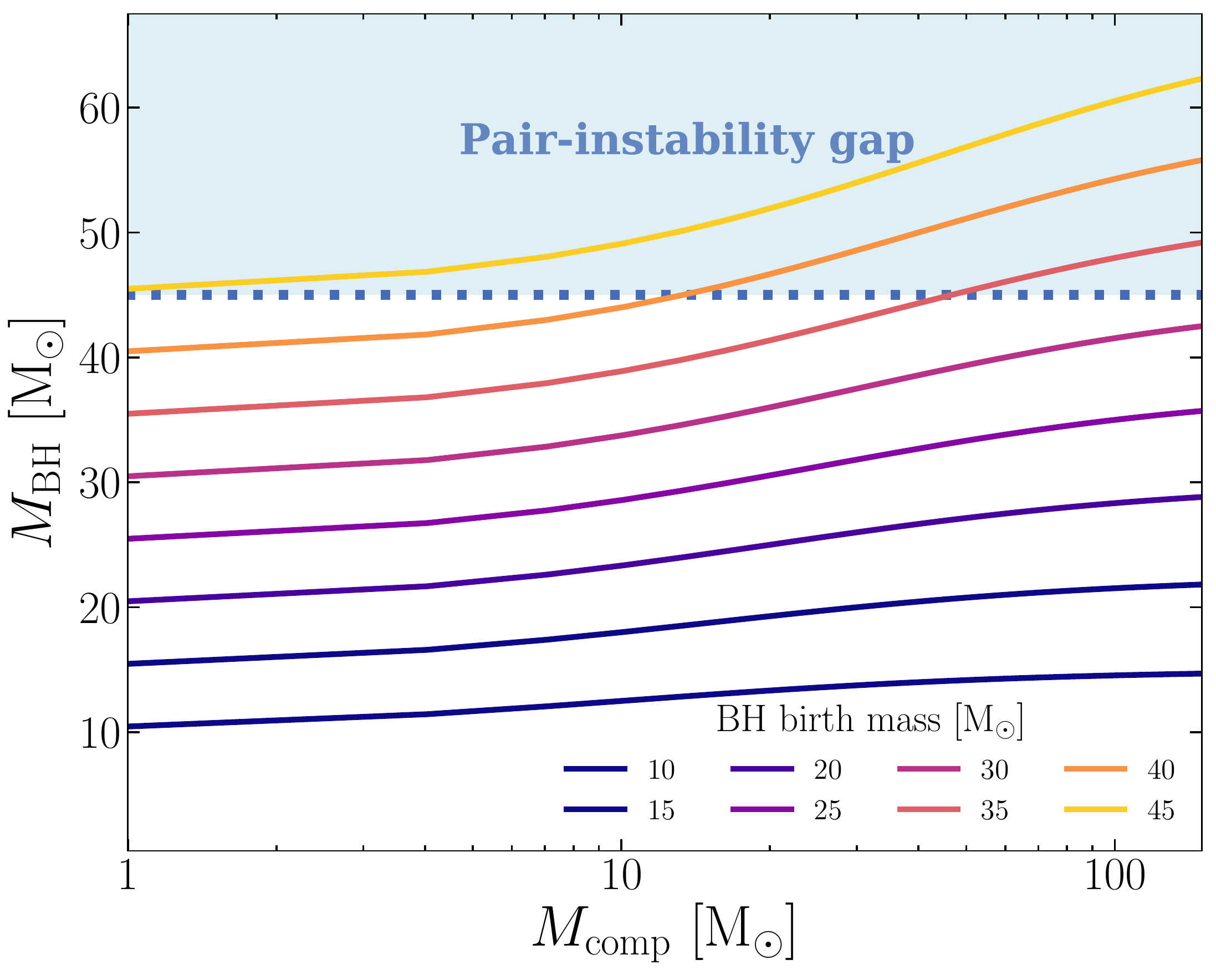}
    \caption{The maximum mass for accreting BHs in model variation two, the CE accretion model, following Equation \ref{eq: BHL accretion rate} as a function of the companion mass at the moment of Roche-lobe overflow. This illustrates that only BHs with a birth mass close to the PISN gap can potentially accrete enough to end with a final mass in the gap. }
    \label{fig:remMassBH}
\end{figure}

The amount that BHs can accrete in this model, $\Delta M_{\rm acc}$, is regulated by Equation~\ref{eq: BHL accretion rate}, which effectively limits the maximum mass that a BH can accrete to about 20\Msun. We illustrate this in Figure \ref{fig:remMassBH}, where we plot $\Mbh =  M_{\rm BH, birth} + \Delta M_{\rm acc}$ as a function of the companion mass at the onset of the CE, for different values of $M_{\rm BH, birth}$. The figure shows that only BHs with high birth masses can produce BHs with masses in the PISN gap. The maximum potential BH mass is about 60\Msun, but this mass can only be achieved under optimal and idealized circumstances. In practice this limit is never reached, both because of sampling effects, and because of stellar winds, which eject part of the envelope before it can be transferred to the BH companion. 
In our numerical simulations, the most massive BH formed is $\Mbhmax \approx 57 \Msun$. 

While this model does not favor the formation of extremely massive systems as found in model variation 1, it does favor the formation of BBHs in tight orbits. A larger number of affected systems can thus merge within a Hubble time through the emission of gravitational waves with respect to the affected systems in model variation 1, as can be seen in the bottom row of Figure~\ref{fig:m1m2}.

\subsubsection{Model variation 3: Combined}
Our combined model variation leads to the combined effects of model variations 1 and 2, the stable accretion and CE accretion models respectively. 
More BBH systems are affected by accretion in the combined model, and the fraction of PISN gap mergers increases. 
However, it does not lead to a significantly larger $\Mbhmax \approx 103$ nor a larger BH mass among the merging BBH population, $\Mbhmax \approx 90\Msun$, than in our second model variation.
We find only about 2\% PISN gap mergers where one of the BHs is more massive than 45\Msun , i.e. $\fgalLIM \approx 2.4\%$.

\subsection{Total mass distribution}
\label{ss total mass dist}

\begin{figure*}[ht]
\centering
  \includegraphics[trim=0.cm 0cm 0cm 0cm, clip, width=0.9\textwidth]{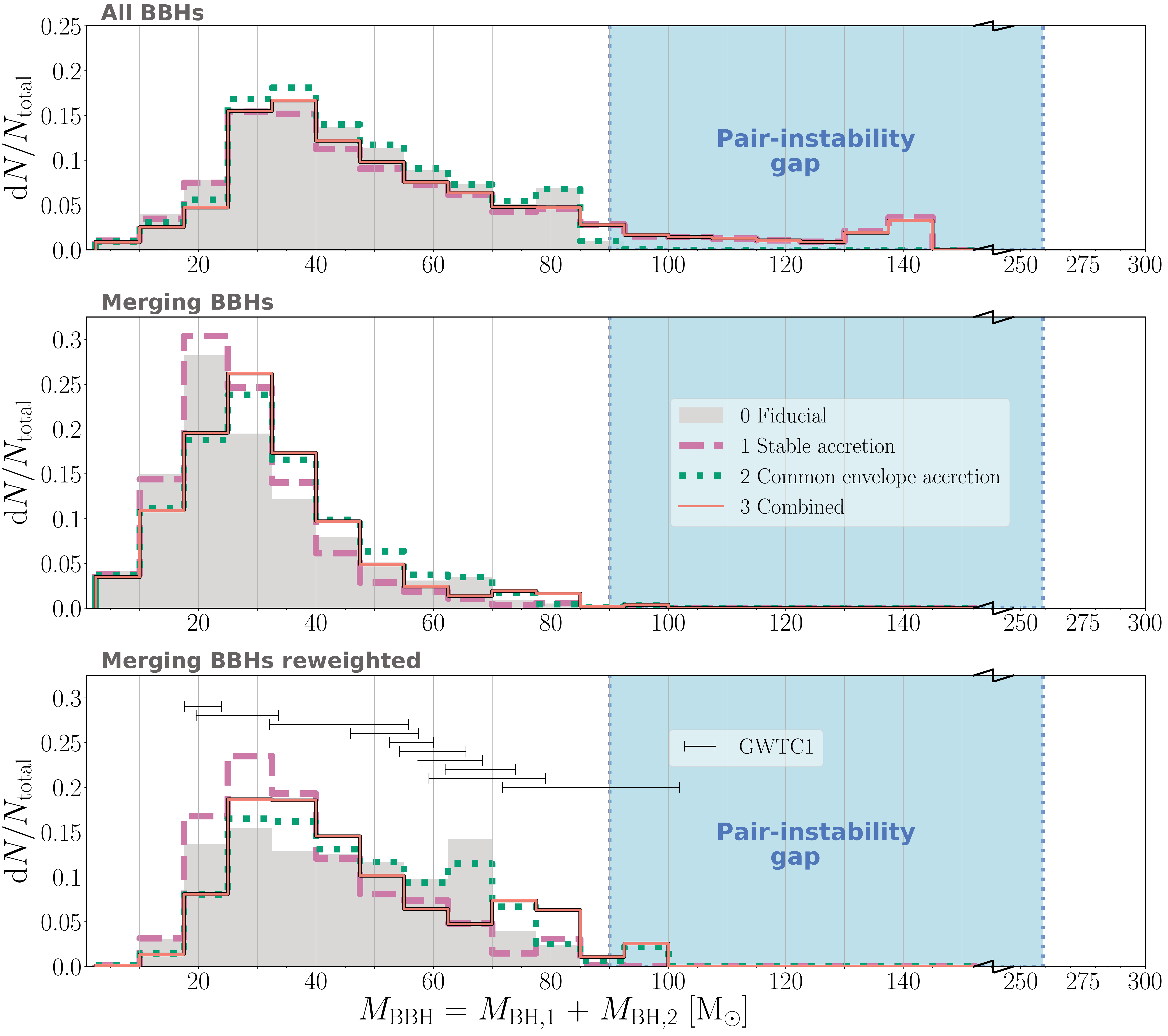}
    \caption{ \textbf{Top:} Weighted distribution of the total BBH masses ($\Mbhtot$) from our fiducial simulation (filled gray), the stable super-Eddington accretion (dashed pink line), the simulation allowing super-Eddington accretion during a CE phase (dotted green line) and the combined model (solid orange line). All distributions are normalized. The light blue region bordered by dotted lines indicates the approximate location of the PISN gap assuming BHs of equal mass. \textbf{Middle:} The same as the top panel, but restricted to BBH systems that are close enough to merge within a Hubble time. \textbf{Bottom:} The same as the middle panel, but the population is re-weighted by the detection bias from LIGO/Virgo \protect\citep{Fishbach2017}. The 90\% confidence intervals of the observed \Mbhtot values from \protect\cite{Abbott2018aGWTC} are also shown as grey horizontal lines, at arbitrary heights.} 
    \label{fig:masshist}
\end{figure*}

\begin{table*}[t]
\caption{The fractions of BBHs with a total mass \Mbhtot higher than 60 and 90\Msun ( \fzestig and \fnegentig, respectively), and the  maximum total BBH mass (\Msysmax ). The errors shown are the 1--$\sigma$ estimate of the statistical sampling uncertainty. }
\label{tab: tot masses}
\begin{tabular}{lccc|ccc}   
  &              \multicolumn {3}{c|}{All} & \multicolumn{3}{c}{Merging}   \\
                                & \fzestig      & \fnegentig    & \Msysmax      & \fzestig    & \fnegentig   & \Msysmax  \\  
Model                           & \%                        &\%             &($\Msun$)      & \%                    & \%                    & ($\Msun$) \\ \hline \hline
  0. Fiducial                   & $23.1\pm0.3$              & $0.0$         & 86            & $5.8\pm0.1$          & $0.0$          & 84         \\
  1. Stable super-Eddington     & $32.2\pm\revision{0.4}$   & $12.7\pm0.2$  & 144           & $2.6\pm0.1$          & $0.01\pm0.00$  & 95        \\%
  2. Common envelope accretion  & $23.4\pm0.3$              & $0.11\pm0.01$ & 99            & $7.1\pm0.1$          & $0.36\pm0.03$  & 99         \\
  3. Combined                   & $32.6\pm0.3$              & $12.2\pm0.2$  & 146           & $6.2\pm0.1$          & $0.45\pm0.03$  & 99        \\
\end{tabular}
\end{table*}

Analysis of gravitational wave merger signals provides a more accurate determination for the total mass of BBH mergers than for the individual BH masses. We therefore show the distribution of the total BBH masses ($\Mbhtot = \Mbheen + \Mbhtwee$) in Figure~\ref{fig:masshist}.
The distributions are normalized and the individual BHs are weighted according to the distributions of their initial parameters. We have checked that the 1--$\sigma$ statistical sampling error is less than 0.001 for all bins. The blue-shaded region marks the location of the PINS mass gap for equal mass systems. 

We furthermore provide the maximum total BBH mass (\Msysmax) and the fractions of BBHs with a total mass \Mbhtot higher than 60\Msun and 90\Msun (\fzestig and \fnegentig respectively, in Table \ref{tab: tot masses}.   

\subsubsection{All BBHs}
The top panel of Figure \ref{fig:masshist} shows the distribution for all BBHs formed in our simulations, including those that are too wide to merge. 

The assumptions we make in the second model variation (super-Eddington accretion during CE) appear to have limited effect on the overall population of BBHs, and the shape of the BBH mass distribution varies little between the fiducial and CE accretion model (models 0 and 2). 

Both the fiducial and CE accretion model avoid the PISN mass gap. For the fiducial model $\Msysmax\approx~86\Msun$ and \fnegentig = 0\%. Although the maximum BBH system mass $\Msysmax\approx 99$ for the CE accretion model, \fnegentig is only 0.1\%.

As discussed in Section \ref{ss comp mass}, the CE accretion model only produces BBH mergers that are marginally in the PISN mass gap. Figure \ref{fig:masshist} shows that such marginal BBH mergers do not stand out as PISN mass gap systems when \Mbhtot is evaluated. 

The stable accretion and combined model variations (models 1 and 3) both display a clear tail of massive systems. The maximum total BBH mass extends to $\Msysmax \approx 144\Msun$ in both models. We find that approximately 12.7\% and 12.2\% of the BBH systems has a mass of $\Mbhtot \geq 90\Msun$ for model variation 1 and 3, respectively.

All model variations peak in $\Mbhtot$ at approximately 30 to 40\Msun.

\subsubsection{Merging BBHs}
The middle panel of Figure~\ref{fig:masshist} only shows BBH systems that merge within a Hubble time. 

The tail of massive systems from the stable accretion and combined model variations is absent in the merging populations, since nearly all of these systems are too wide to merge (as discussed in Section~\ref{ss comp mass}). 

We see that the subset of merging BBHs from the first model variation (super-Eddington accretion during stable mass transfer) is not significantly different from the fiducial population of merging BBHs. For both the fiducial and the first model variation $\fnegentig = 0\%$ and $\Msysmax \approx 84\Msun$.

The second model variation (super-Eddington accretion during CE) affects the subset of BBHs that merges more strongly than the first model variation. We see that the peak of the merging distribution is shifted to higher masses with respect to the fiducial population, which is reflected in the higher value of $\fzestig \approx 7.1\%$. 

In other words, mass distribution of merging BBHs is shifted to higher masses in the CE accretion model with respect to the fiducial distribution.
This effect is also visible in the combined model variation, which closely follows the merging BBH distribution of the CE accretion model. 

Although the peak of the merging distribution is shifted to higher masses for our second and third model variations, they do not produce a significant amount of PISN mass gap events in terms of \Mbhtot. Only about $0.36\%$ and $0.45\%$ of the merging populations has a $\Mbhtot >90\Msun$ for model variation 2 and 3. 

\subsubsection{Merging BBHs reweighted}
In the bottom panel of Figure~\ref{fig:masshist} we apply a simple re-weighting to the merging distribution to account for the detection probability which scales approximately as $(\Mbheen)^{2.2}$, following \cite{Fishbach2017}. 
\revision{In this panel, we show the 90\% confidence intervals of the observed \Mbhtot values from \protect\cite{Abbott2018aGWTC} at arbitrary heights with horizontal grey lines. This shows that the total mass distribution as produced by our fiducial model is able to form BBHs with total masses similar to the detections from LIGO and Virgo's first and second observing runs.} 

Re-weighting the distribution results in the largest deviations from the fiducial simulation.  
In general the re-weighting has a flattening effect on the distribution of BBH masses, since the massive end ($\Mbhtot >60\Msun$) of the distribution is boosted, while the intrinsic BBH distributions peak at low masses (approximately between 20 and 30\Msun).

The small fraction of BBHs with masses in the PISN mass gap from the second and third model variation becomes visible due to the re-weighting.
However, we see that BBH mergers with masses in the range $90\Msun\leq\Mbhtot<100\Msun$ only constitute a few percent of the re-weighted distribution. Moreover, none of the model variations produces a merging BBH system with a mass of $\Mbhtot \geq 100\Msun$.

We conclude that, despite our extreme assumptions regarding accretion onto BHs, none of our model variations is able to significantly populate the PISN mass gap with systems that merge in a Hubble time. Under our most extreme assumptions, we find that in only about $0.45\%$ of all cases, the BBH mass \Mbhtot exceeds 90\Msun. 

\subsection{\revision{Mass ratios} }
\label{ss results mass ratio}

\begin{figure*}[ht]
\centering
  \includegraphics[ width=\textwidth]{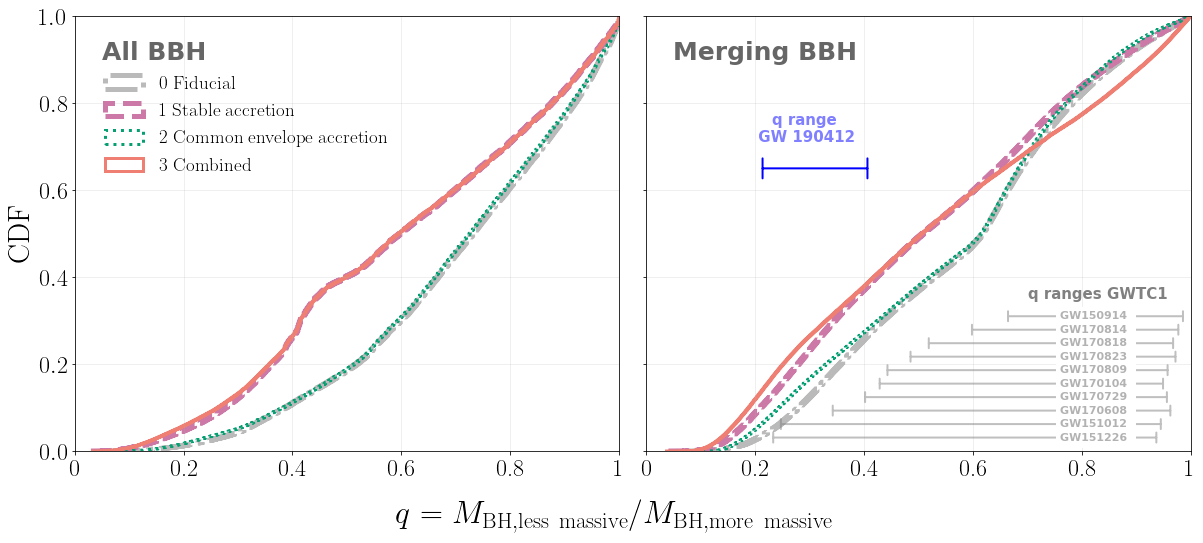}
    \caption{\revision{Cumulative distribution function of the mass ratio $q$, defined as the ratio of the less massive over the more massive BH. The left panel shows all BBHs, while the right panels displays only the BBH population that merges within a Hubble time. The 90\% confidence interval of observed mass ratios are shown as gray lines for observations from LIGO/Virgo's first and second observing run \protect\citep{Abbott2018aGWTC}, and as a blue line for GW190412. The CE accretion model has little effect on the mass ratio distribution of BBHs. The stable accretion model variation and combined model variation shift the mass ratios to lower values.} }
    \label{fig:CFFq}
\end{figure*}

\revision{Figure \ref{fig:CFFq} displays the cumulative distribution function of the mass ratio, $q$, defined as the ratio of the less massive over the more massive BH.  
Figure \ref{fig:CFFq} shows that the CE accretion model leads to similar mass ratios as the fiducial model, but results in a slightly larger fraction of mass ratios with $q \leq 0.35$ when considering the merging population (dash-dotted gray, and dotted green lines in Figure \ref{fig:CFFq}). 

The stable accretion model and combined model (dashed pink, and solid orange line in Figure \ref{fig:CFFq}) lead to a higher fraction of low mass ratio systems than the fiducial model. For the combined model, we find that about 40\% (50\%) of all (merging) BBHs have a mass ratio of $q\leq0.5$. For the fiducial model, we find that about 20\% (40\%) of all (merging) BBHs have a mass ratio of $q\leq0.5$. Moreover, the stable accretion model and combined model allow for more extreme mass ratios with respect to the fiducial population, down to $q \approx0.1$.
These low mass ratios are caused by accretion onto the first-born BH in the models where we allow for super-Eddington accretion. This accretion increases the mass of the first-born BH, and thus leads to BBHs with lower mass ratios. 

The 90\% confidence interval of observed mass ratios from LIGO/Virgo's first and second observing run are shown at the bottom right of Figure \ref{fig:CFFq} \citep{Abbott+2016_properties}. For most detections, the mass ratios are relatively poorly constrained. Of special interest is the recently announced GW190412, which is the only system published so far with significantly unequal masses,  \citep[$q = 0.28^{+0.13}_{-0.07}$, ][blue line in Figure \ref{fig:CFFq}]{Abbott2020gw190412}.  The formation of systems with such mass ratios is more common in our models that allow for super Eddington accretion. We find that about 30\% of the merging fiducial population has a mass ratio of $q \leq 0.41$, which is the upper limit of the 90\% confidence interval for the mass ratio of GW190412. This fraction increases to 40\% of the merging BBH population for the combined model variation.

If we consider systems for which the individual masses coincide with the the component masses inferred for GW190412 \citep[\Mbheen=29.7$^{+5.0}_{-5.3}$\Msun, \Mbhtwee~=8.4$^{+1.7}_{-1.0}$\Msun,][]{Abbott2020gw190412}, we find that 0.5\% of the merging fiducial population coincides with GW190412. This fraction increases to 3.6\% for the merging population of the Combined model.  We further note that GW190412 shows evidence for spin, which may be expected for the accreting BH. However, see Section \ref{ss : compare spins} for a more in depth discussion of the spins. 

}

\subsection{BBH merger rates}
\label{ss merger rates}

\begin{table*}[t]
\caption{Estimate of the BBH merger rates for our chosen fixed metallicity. We provide the total number of BBHs formed per unit star-forming mass that merge within a Hubble time, $\rm dN_{t_H}/dM_{SFR}$, and within 10\Gyr, $\rm N_{10}/M_{SFR}$, respectively. The BBH merger rate for a synthetic Milky Way-like galaxy (\Rmwg), the volumetric BBH merger rate \Rvol, and the volumetric merger rate for BBHs with a component $\Mbh > 45 \Msun$. The 1--$\sigma$ estimate of the statistical sampling uncertainty is shown whenever it exceeds 0.5\% of the relevant value. }
\label{tab: merger rates}
\begin{center}

\begin{tabular}{lccccc}
                                & $\rm N_{t_H}/M_{SFR}$          & $\rm N_{10}/M_{SFR}$          & \Rmwg                & \Rvol                 & \RvolLIM                  \\
   Model                        & ($\Msun \cdot$ 10$^{-5}$)    & ($\Msun \cdot$ 10$^{-5}$)      & (Myr$^{-1}$)          & (Gpc$^{-3}$yr$^{-1}$) & (Gpc$^{-3}$yr$^{-1}$)     \\ \hline\hline
 0. Fiducial                    & $2.3$                        & $2.2$                        & $77\pm\revision{0.3}$              & \revision{$897\pm4$}                 & $0.0$                     \\
 1. Stable super-Eddington      & $2.4$                        & $2.3$                        & $81\pm\revision{0.3}$              & \revision{$937\pm4$}                 & $1.2$                     \\
 2. Common envelope accretion   & $2.2$                        & $2.0$                        & $72\pm\revision{0.3}$              & \revision{$832\pm3$}                 & $11$                      \\
 3. Combined                    & $2.1$                        & $2.0$                        & $71\pm\revision{0.3}$              & \revision{$825\pm3$}                 & $19$           
\end{tabular}
\end{center}
\end{table*}

We briefly discuss simple estimates for the merger rates that can be obtained directly from our simulations. For this we closely follow the procedure outlined by \cite{Dominik+2012}, detailed in Appendix~\ref{app: merger rate}. Table~\ref{tab: merger rates} provides an overview of the quantities discussed here. 

In this work, we consider simulations at a fixed metallicity of Z=0.001, representative for the low metallicity environments in which heavy BBHs are believed to form \citep[e.g.][]{Abbott+2016_astrophysical_implications}. By choosing a fixed low metallicity, we overestimate the BBH formation rate and BBH merger rate. 
\revision{We furthermore assume the optimistic CE model as discussed in \cite{belczynski2017evolutionary}. The optimistic CE model tends to lead to an overprediction of the BBH merger rate \citep[e.g.][]{Dominik+2012}. \cite{Stevenson+2017} find merger rates which are approximately 3 times lower for the pessimistic model with respect to the optimistic model at $Z=0.001$. Assuming the optimistic model, their BBH merger rates at $Z=0.001$ are comparable to the rate estimate we find for our Fiducial model.} 

Our estimates of the BBH merger rates should be considered as rough upper limits that enable comparison to other work. 
For a more careful treatment we refer to \cite{Neijssel+2019} \revision{and \cite{Stevenson+2019},} who consider different metallicities and account for the abundance evolution through cosmic time. They also \revision{both} use the \COMPAS suite, while assuming initial conditions that are very similar to those adopted in our fiducial model. They indeed find merger rates that  are consistent with the rates from \cite{Abbott2018aGWTC}. 

We first estimate the number of BBHs formed per unit star forming mass that merge within a Hubble time, $\rm dN_{t_H}/dM_{SFR}$ and the number of BBHs formed per unit star forming mass that merge within 10\Gyr $\rm dN_{10}/dM_{SFR}$. We find that both quantities vary only slightly across our model variations, as can be seen in Table~\ref{tab: merger rates}.  
The physical assumptions we varied primarily affect the amount of mass that BHs accrete and therefore their masses, but these assumptions do not significantly affect the rate of BBH mergers. 

We also estimate the total merger rates for a synthetic Milky Way-like galaxy, \Rmwg and the consequential total volumetric merger rates \Rvol. 
As stated above, these rates are higher than the current estimates from \cite{Abbott2018aGWTC} due to the fixed low metallicity, however they are consistent with estimates from \cite{Stevenson+2017} and \cite{de-Mink+2015}.
\Rmwg (and consequently \Rvol) are comparable for the fiducial and first model variation. However, they are slightly lower for the second and third model variation, which implies that allowing for accretion during a CE leads to slightly fewer BBH mergers. This can be understood as our CE accretion model leading to more `failed' common envelopes that end in a stellar merger instead of a BBH.

Applying the fractional rates, $\rm f_{\rm 45}$, to the volumetric merger rate for BBHs (\Rvol), results in our estimates of the PISN mass gap event rate, \RvolLIM. 
For our first model variation, the estimates of the PISN mass-gap-event rates \RvolLIM,  are consistent with the rates inferred by \cite{Fishbach2019} for gravitational wave events in the first and second observing run. They find $\RvolLIM=3.02^{+12.97}_{-2.28}$Gpc$^{-3}$ yr$^{-1}$ under the assumption of a flat-in-log prior for the mean merger rate per bin. Assuming a power-law prior, \citet{Fishbach2019} constrain the PISN mass gap merger rates to $\RvolLIM = 1.79^{+2.30}_{-1.23}$ Gpc$^{-3}$ yr$^{-1}$. 
The second and third model variation lead to estimates of the PISN mass-gap-event rates that are higher than the current estimates from \cite{Fishbach2019}.

The uncertainties quoted in Table \ref{tab: merger rates} result from the sampling procedure. We note that the model uncertainties are much larger, by orders of magnitude \citep[e.g.\ ][]{Dominik+2012, de-Mink+2015,Chruslinska+2018} and that the rates are affected by the choice of a constant metallicity $Z = 0.001$.


\section{Distinguishing between different pathways to PISN mass gap events}
\label{sec: result compare path}
In this paper we consider the possibility to form PISN mass gap events through the isolated binary evolutionary channel. Various other pathways have been proposed to produce PISN mass gap events, see Section~\ref{sec:intro pathways} for a brief overview. In this Section, we compare our results to the findings for other pathways that have been proposed in the literature to create PISN mass gap events. 

This comparison is not straightforward. Different studies have adopted different input assumptions, for example concerning the location of the PISN mass gap. Moreover, the few quantitative predictions that exist to date often rely on relatively crude assumptions for complex physical processes. It is to be expected that these predictions will change with time as the models become more sophisticated. The comparison we present here reflects what is available in the literature to date.  

\begin{figure*}[ht]
    \centering
    \includegraphics[width=0.8\textwidth]{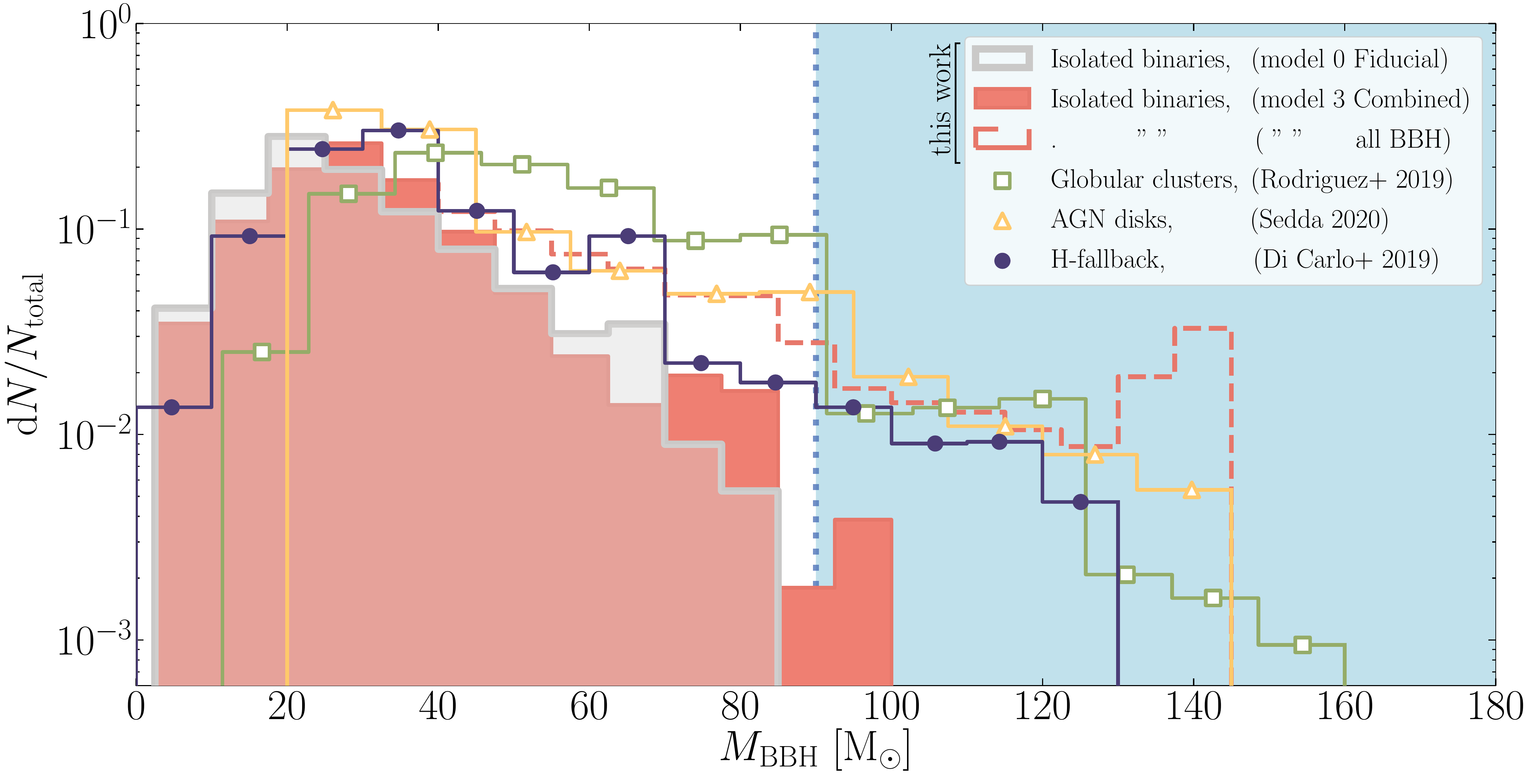}
    \caption{\Mbhtot distribution for BBH systems from different pathways into the PISN mass gap. We compare the distribution of total BBH masses, $\Mbhtot$, of multiple generations of BH-mergers in globular clusters \protect\citep[green open squares,][]{Rodriguez2019}, multiple generations of BH-mergers in AGN disks \protect\citep[yellow open triangles,][]{sedda2020birth} and stellar mergers in young stellar clusters assuming full H-fallback during BH formation \protect\citep[closed dark blue circles,][]{DiCarlo2019a} to the effect of super-Eddington accretion in isolated binaries as discussed in this work. The grey line displays the merging population of BBH systems from our Fiducial model (model 0) The dashed and filled orange distributions respectively show the total and merging population of BBH systems from our most optimistic model variation (model 3 combined). The light blue region bordered by dotted lines indicates the approximate location of the PISN gap assuming BHs of equal mass. All distributions are normalized to their respective population of merging BBHs. This is therefore \textit{not} a prediction for the number of PISN mass gap events.}
    \label{fig: compare to others}
\end{figure*}

We first briefly compare different pathways and their predictions for the shape of the BBH mass distribution (Section~\ref{ss : compare dist} and Figure~\ref{fig: compare to others} ), followed by predictions for the maximum masses  (Sect.~\ref{ss : compare max m}), mass ratios (Section~\ref{ss : compare q}) and BH spins (Section~\ref{ss : compare spins}). 
A schematic overview is provided in Figure \ref{fig:cartoon and table}.

\subsection{The shape of the BBH mass distribution \label{ss : compare dist}}
\label{ss:BBH tot mass dist}

In Figure~\ref{fig: compare to others}, we compare various predictions for the distribution of total BBH masses, \Mbhtot, for systems that merge within a Hubble time. This figure is similar to the middle panel of Figure~\ref{fig:masshist} except for the use of a logarithmic y-axis, to highlight the differences in the tail of the distribution. We consider BBHs with total masses above 90\Msun to be PISN mass gap events.  All distributions are normalized to their respective population of BBHs. This figure thus allows us to compare the shape of the distributions but, at present, cannot be used to infer which pathway contributes most significantly to PISN mass gap since the relative contribution from each pathway is unknown.

\cite{sedda2020fingerprints} argue that with enough detections, the fraction of mergers with masses falling in the low or high mass end of the BBH mass distribution can be used to place constraints on the relative contribution of different pathways to the overall population.

\subsubsection{Isolated binaries} 
Our most optimistic simulation for super-Eddington accretion in isolated binaries, which allows for accretion onto BHs during both stable mass transfer, and CE events is shown in filled red in Figure~\ref{fig:masshist}. 

We find that isolated binaries will contribute less than $0.45\%$ of mergers with $\Mbhtot \geq 90\Msun$.
As discussed in Section \ref{sec:results}, although the maximum value of \Mbhtot, i.e. \Msysmax, increases with respect to our fiducial simulation (model 0, shown in grey for reference), isolated binary evolutionary does not significantly populate the PISN mass gap. 

Our results show that super-Eddington accretion in interacting binaries can produce BBHs with total masses larger than 90\Msun, but due to their typically large orbital periods (of approximately 100 days), we do not expect these systems to merge as a result of isolated binary interaction alone. 

It is conceivable that a fraction of these binaries experience a decay of their orbits due to external factors.

For example, a significant number of massive binary systems are born as part of a triple system or even higher order multiple \citep[e.g.][]{Sana+2014}. Secular interactions with a third companion such as Lidov-Kozai cycles have been proposed to increase the merger rate \citep{Toonen2016, Kimpson2016,Antonini2017}.

Furthermore, a fraction of massive binaries is born in dense stellar environments, such as a globular or nuclear star cluster. These binaries and their BH remnants can be affected by a sequence of dynamical encounters and exchanges \citep{Rodriguez+2016}. The heavy BHs in our models are good candidates for dynamical interactions since they will be among the most massive BHs formed in the cluster. As these sink to the center of the cluster they will be prone to interact and take part in dynamically-assisted mergers. 

Simulating the combined effects of super-Eddington accretion and external effects is beyond the scope of this paper. However, we can explore the upper limits of the contribution from super-Eddington accretion by considering the distribution of total masses for our full BBH population including those that are too wide to merge due to gravitational waves alone (model 3 ``al'', dashed red line in Figure \ref{fig: compare to others}).  
About 12\% of the BBHs in this distribution have a total mass \Mbhtot, that exceeds $2\times 45 = 90\Msun$. This should be considered as an extreme upper limit that we do not believe to be realistic. The distribution of all BBH systems in our combined model extends up to 145\Msun, showing a rise around 130--145\Msun. This pile up results from BBH systems where the first born BH gained mass through stable accretion from its companion. This distinguishes this distribution from other pathways, which all decline at high masses.

\subsubsection{Globular clusters }  We show results for globular clusters from \citealt{Rodriguez2019} (green open squares). In their simulations, close BBHs form and tighten as a result of dynamical interactions in the dense core of a globular star cluster. Massive BHs that form as the result of an earlier BBH merger may stay bound to the cluster if the merger recoil is sufficiently small. These so-called second generation BHs can have masses in the PISN gap. They can give rise to PISN mass gap events if they pair and merge with a third BH.  We show the results from the most optimistic model for all redshifts by \citealt{Rodriguez2019} (as shown in the top left panel of their figure~3). This model assumes zero birth spin for all black holes. Low birth spin minimizes the BBH merger recoil during the first merger event, enabling the resulting BH to take part in a second merger.

Their distribution displays a significant drop around 90\Msun, which corresponds to two times the maximum value for first generation mergers. For total masses between 90 and 125\Msun the distribution is dominated by $\rm 1^{st} + 2^{nd}$ generation mergers.
A second drop exist, near 125\Msun, which is close to three times \Mbhmax. Events with masses in excess of 125\Msun are primarily the result of $\rm 2^{nd} + 2^{nd}$ generation BHs, which very rare. Their distribution extends to $\Mbhtot \approx 150\Msun$. 

\cite{Rodriguez2019} find that about $4\%$ of the detected BBHs will have $\Mbhtot \geq 100\Msun$ for their most optimistic model assuming zero birth spin for all BHs.
This pathway is the most efficient at producing very massive events among all those we consider, at least in relative terms. The fraction of events where the total mass exceeds 90\Msun is about 5\% in their simulations. However, when a less optimistic model is assumed, i.e. when the birth spin for BHs is assumed to be non-zero, the rate of PISN mass gap events drops significantly. For example, when assuming a birth spin of $\chi_{\rm birth} = 0.5$ for all BHs, they find that less than 0.1\% of all BBH mergers will have a total mass $\Mbhtot\geq 100\Msun$ \citep[as shown in the bottom left panel of figure~3 from][]{Rodriguez2019}.

\subsubsection{AGN disks} AGN disks have been proposed as promising sites that allow for a sequence of multiple mergers and thus the creation of PISN gap mergers in a similar way as globular clusters. The difference between these two pathways arises from the larger escape speeds in AGN disks due to a deeper potential well. This opens the possibility for higher generations of BHs in AGN disks, while the contribution from $\rm 2^{nd} + 2^{nd}$ and $\rm > 3^{rd}$ generation BHs is expected to be negligible in globular clusters \citep{GerosaBerti2019}. 

We compare our mass distribution to predictions from \cite{sedda2020birth}, extracted from the top panel of their figure~19, shown as yellow open triangles in Fig. \ref{fig: compare to others}. Their distribution extends to about 140\Msun. 

The fraction of events where the total mass exceeds $\Mbhtot > 2 \times 45 = 90\Msun$ is about 4\% in their simulations, which is comparable to the predictions by \citet{Rodriguez2019} for globular clusters when BHs are born with zero spin. 

\subsubsection{Fallback of a hydrogen-rich envelope }
Finally, we compare to a pathway studied by \citet{DiCarlo2019a}. They consider stellar mergers occuring in young star clusters, involving at least one evolved star. Such mergers are poorly understood, but they may produce stars with overmassive hydrogen envelopes and relatively small cores \citep{Vigna-Gomez+2019}. If the core mass of such a merger product is below the limit for pair pulsations, it is expected to collapse directly to a BH. If it is assumed that the massive hydrogen envelope is entirely accreted onto the forming BH, this can result in a BH with a mass in the PISN mass gap. Dynamical interactions within the cluster could later pair such a BH with another BH, possibly facilitating a BBH merger. 

We compare to the simulations from \citet[][their Figure~5 ]{DiCarlo2019a}, shown as closed dark blue circles in Figure \ref{fig: compare to others}. They find BBH mergers with total masses up about 130\Msun. This is higher than what we expect from isolated binaries, but lower than what is claimed for globular clusters and AGN disks. 

The fraction of events where the total mass exceeds $\Mbhtot > 2 \times45 = 90\Msun$ is about 3\% in their simulations, which is also slightly below the predictions for globular clusters and AGN disks.

\subsection{Predictions for masses, mass ratios and spins}
\label{ss: observables}
The overview of predicted BBH mass distributions as displayed in Figure \ref{fig: compare to others} show that the BBH population cannot be explained by the isolated binary evolution channel alone if more than $1\%$ of all BBH mergers has a mass higher than $90\Msun$. We will now discuss other observables that might help distinguish between the different pathways considered in this work.

\begin{figure*}[ht]
   \centering
    \includegraphics[trim=1.cm 3.8cm 1cm 4.85cm, clip, width=\textwidth]{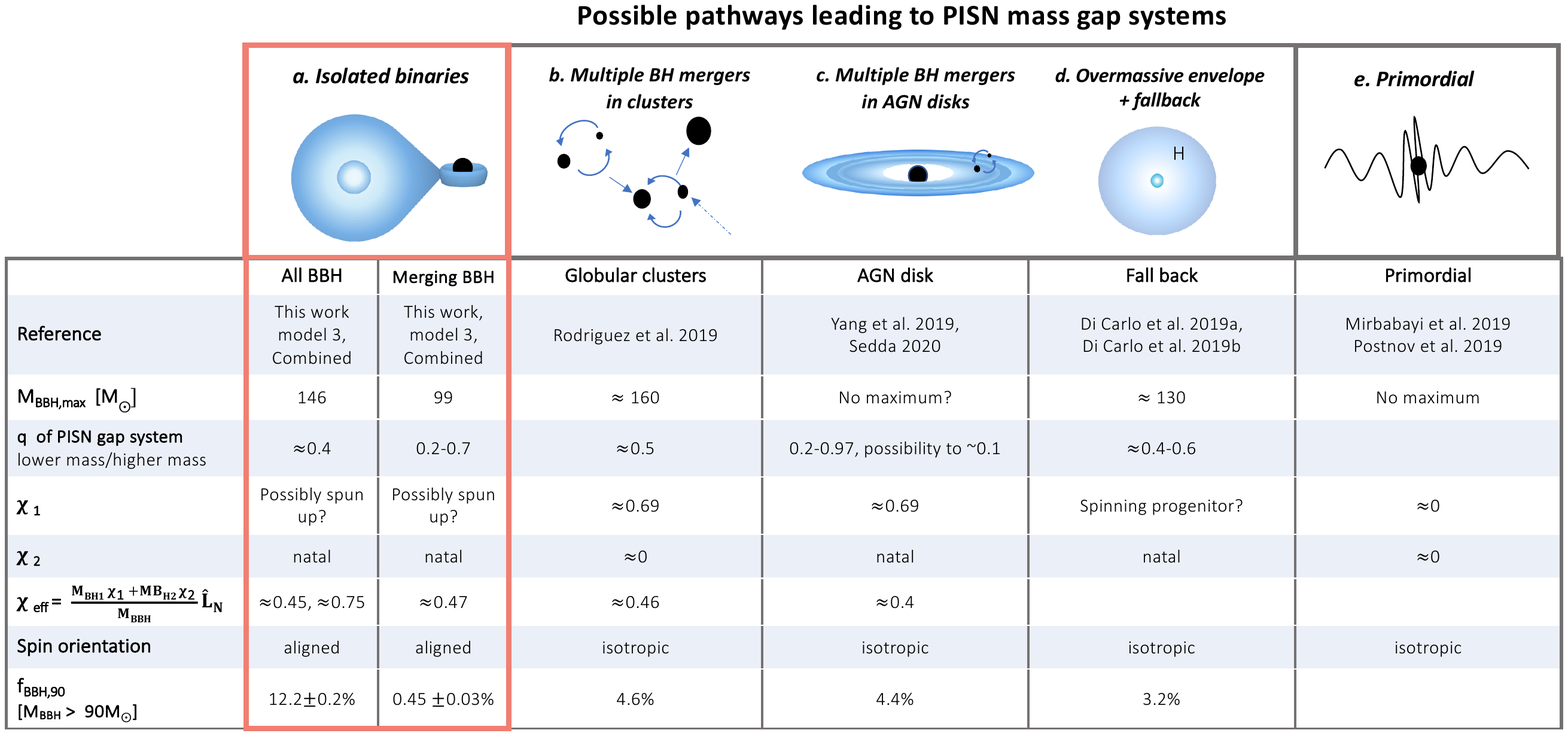} 
    \caption{Predicted characteristics of PISN mass gap events from different pathways. We compare the maximum BBH mass, \Msysmax, the mass ratio $q$ of the system defined as the ratio less massive over the more massive BH mass, the dimensionless spin parameters of the BHs $\chi_1$ and $\chi_2$, the effective spin parameter, \chieff, the \revision{expected} spin orientation and the fraction of BH systems with $\Mbhtot > 90\Msun$. }
    \label{fig:cartoon and table}
\end{figure*}

\subsubsection{Maximum masses} \label{ss : compare max m}
 In all our model variations, the amount of mass that a BH can accrete is ultimately capped by the envelope mass of the donor star. The most massive BBH system that merges within a Hubble time has a mass 99\Msun, though this is for extreme assumptions. We do find more massive systems, up to about 144\Msun, but for those we would need to invoke an external mechanism to merge the system. 
 
 This maximum mass distinguishes this channel from predictions by other proposed pathways. Simulations of multiple BH mergers in massive globular clusters and AGN disks are found to result in maximum BBH masses of about $160\Msun$ \citep{Rodriguez2019} and about $145\Msun$ \citep{sedda2020birth} respectively. However, as long as the merger product remains bound to the merger environment, there is no reason to believe that these pathways have to adhere to any maximum value of \Mbhtot. 
 
\cite{DiCarlo2019a} find a maximum of 130 \Msun for a metallicity of Z = 0.02. The maximum mass resulting from PISN mass gap events as discussed in \cite{DiCarlo2019a} and \cite{DiCarlo2019b} are in essence the sum of the maximum BH mass and the envelope mass of the progenitor at the moment of BH formation (i.e. $\Mbhmax +\rm M_{env}$). When complete fallback of the overmassive envelope is assumed, the maximum BBH system mass is thus capped by the maximum possible envelope mass prior to BH formation.  \\


\subsubsection{Mass ratios} \label{ss : compare q}
\revision{PISN mass gap events resulting from the classical binary evolutionary channel may also be distinguished by their mass ratios. Here we define $q$ to be the mass ratio of the less massive over the more massive BH.  As discussed in Section \ref{ss results mass ratio}, accretion onto the first born BH increases its mass, leading to events with more extreme mass ratios. Figure \ref{fig:CFFq} displays the mass ratios for all BBH systems. When we focus on the mass ratios of PISN mass gap systems, we find that our Combined model predicts BBH systems that peak strongly at $q \approx 0.4$. }

\revision{The merging population of PISN mass gap systems from our Combined model predicts a wider range of mass ratios: $0.2 < q < 0.7$.}
These mass ratios are comparable but slightly lower than those for PISN gap events from stellar and globular clusters, which are expected to peak around $q\sim 0.4-0.6$ and $q\sim0.5$, respectively \citep{Rodriguez2019,DiCarlo2019a}. 
 
Mass ratios from BBH mergers in AGN disks are still highly uncertain, but  \cite{mckernan2020black} predict that the median mass ratios of BBH mergers in AGN disks will range from 0.20 to 0.97. 
However, when a higher generation BH is involved (i.e. a BH formed through multiple consecutive BBH mergers), mass ratios can be expected to drop to very small values of $q$. This could possibly push $q$ down to values lower than $q \sim 0.1$ when around 10 consecutive mergers or more are allowed.

\subsubsection{Spins} \label{ss : compare spins}
\citet{Bardeen1970} argues that spin-up is expected as a result of accretion in the case of a thin disk  \citep[see also][]{kingkolb1999, OShaughnessy+2008}. However, the (significantly) super-Eddington accretion rates considered in this work are expected to result in thick accretion disks or even near-radial inflow \citep{Volonteri+2005, Begelman+2006, Pezzulli+2016,Johnson+2016}. It is not clear whether this accretion geometry will lead to significant spin up. For example, \cite{Tchekhovskoy2012} show that accumulation of magnetic flux around the central regions of the accreting BHs might cause the BH to spin down instead of spinning up.  Therefore we cannot, at present, confidently predict the final spins of BHs.

Nevertheless, we attempt to provide an upper limit using the expression in Eq. 4 of \cite{Bardeen1970} for thin accretion disks. Using this and the assumption of zero natal spins, we find that the vast majority ($>75\%$) of the accreting BHs are spun up to the maximum possible spin value in our first model variation for stable mass transfer. This results in effective spins of $\rm 0.65 < \chieff < 0.8$.
In our second model variation for CE accretion, we find effective spins of $\rm 0.2 < \chieff < 0.6$.
Our combined model spans the whole range of effective spins and results in $\rm 0.2 < \chieff < 0.85$, with two distinct peaks, one around $\chieff\approx 0.45$ and one around $\chieff \approx 0.75$.
The merging population of the combined model is dominated by the effective spins of the CE accretion model and spans the range $\chieff\approx0.2$--0.6, with a peak around $\chieff \approx 0.47$.

We expect PISN mass gap events created through super-Eddington accretion in isolated binaries to result in relative alignment of the BH-spin with the orbit. The spin of the first-born BH will likely align with the orbit during the mass transfer phase. However, \revision{a} natal kick of the second born BH \revision{could possibly} tilt the orbit. Given the uncertainties in the spin itself we have chosen not to model this\revision{, but we expect no, or very low velocity, natal kicks for the most massive BHs}. 

For globular clusters, the optimal conditions for PISN mass gap events as discussed in \cite{Rodriguez2019} require two non-spinning BH for the first generation of BHs, which are expected to produce BHs with spins strongly peaked at $\rm \chi \approx 0.69$. This implies that one of the BHs involved in the PISN mass gap event is expected to have a spin of $\rm \chi_1 \approx 0.69$, while its companion is expected to have its natal spin of $\rm \chi_2 \approx0$.

\cite{Yang2019} predict that the effective spin distribution of BBHs is dominated by $\rm1^{st} + 2^{nd}$ generation BHs. This suggests that the spin of the incoming second generation BH will be strongly peaked around $\rm \chi_1 \approx 0.69$, as explained above. Assuming all first generation BHs have the same mass, they find that due to the random alignment of spins, this results in an effective spin distribution that is peaked strongly around $\rm \chi_{eff} \approx 0.4$.

Furthermore, it is highly uncertain what BH spin is expected for BBHs in stellar clusters when the BH-progenitor is the end product of a stellar merger \citep[as suggested in][]{DiCarlo2019a}. One might intuitively argue that such a stellar merger would lead to a spinning BH progenitor, however \cite{Schneider_2019} show that stellar mergers can result in a slowly spinning merger product.  
In any case, the connection between BH spin and its progenitor is highly uncertain, even more so when an overmassive envelope is speculated to fall back during BH formation \citep[see e.g. ][]{Heger+2005,Lovegrove+2013, belczynski2017evolutionary}. 

The BBH systems leading to PISN mass gap events formed in globular clusters, AGN disks, and stellar clusters are all dynamically assembled.
For such dynamically-assembled BBHs, the angle between the orbital angular momentum and the BH spins is expected to be distributed isotropically \citep{Rodriguez_2016_spin}. This results in misaligned spin orientation in the majority of cases and suggests that the distribution of $\chi_{eff}$ is also symmetric and centered on zero \citep{Rodriguez2019}.

Lastly, the spin of primordial BHs is conventionally believed to be small \citep{mirbabayi2019spin,Luca_2019}. BBHs consisting of primordial BHs are expected to be dynamically assembled and thus have isotropically distributed spins \citep{Rodriguez_2016_spin}.\\

\revision{In conclusion, it is extremely difficult to distinguish between the different pathways to PISN mass gap events. At the time of writing, predictions from different pathways for the maximum masses, mass ratios, and BH spins are not sufficiently constrained to decisively differentiate between pathways. 

In light of the above discussions, we find that the combination of extreme mass ratios and an aligned spin orientation in a BBH system with $\Mbhtot \leq 100 \Msun$ could be indicative of BHs that underwent super-Eddington accretion from a companion star (Figure~\ref{fig:cartoon and table}). 
We expect that our ability to distinguish between the different pathways will be improved with upcoming gravitational-wave surveys which will enhance constraints on both the rates and properties of (PISN mass gap) mergers. }

\section{Discussion} 
\label{sec:discussion}
This work examines whether the classical isolated binary evolutionary channel can produce BBH mergers with a component in the pair instability mass gap (a PISN mass gap event).  Under our most extreme assumptions (i.e., those that favor the mass growth of black holes most strongly) we find about $2\%$ of all BBH mergers at $Z = 0.001$ to be PISN mass gap events and we find a maximum mass for a BH involved in a PISN mass gap event of $\Mbhmax = 90\Msun$. Moreover, under these assumptions, we find only about $0.45\%$ of the merging BBH systems have a total mass $\Mbhtot \geq 90\Msun$, and we find no merging BBH systems with masses of $\Mbhtot \geq 100\Msun$.

We discuss how robust these main findings are against variations in the assumptions about mass transfer in Sections~\ref{ss: shrinking the orbits} and \ref{ss: accrete even more}, and consider further caveats in Section~\ref{ss: caveats}. We conclude by speculating on the effects of super-Eddington accretion in binaries to the lower mass gap in Section \ref{ss: lower mass gap}.

\subsection{Variations in mass transfer}
\label{ss: var in MT}
Most of the heavy BBHs in our simulations are too wide to merge within a Hubble time (see Section \ref{sec:results}). The BBHs that do merge as PISN mass gap events are only marginally in the PISN mass gap since they have accreted less mass than their non-merging counterparts.
We first discuss additional mechanisms that could shrink the BBH orbits and whether this could increase the rate of PISN mass gap events in Section~\ref{ss: shrinking the orbits}. This is followed by a discussion on whether there are any possibilities left to accrete more mass onto the BHs in Section \ref{ss: accrete even more}.

\subsubsection{Shrinking the binary orbit}
\label{ss: shrinking the orbits}
In our first model variation, we have assumed that BHs can accept all the mass that is available from their donor star. We thus assume that the total mass and angular momentum is conserved. If instead a fraction of the mass is lost, this lost mass will carry away angular momentum, which leads to a different orbital evolution.

Observations of X-ray binary systems show evidence for outflows resulting from the accretion disk around a black hole \citep[e.g.][]{Blundell2005,Remillard+2006}. The effect of such an outflow on the binary orbit can be modeled assuming that a fraction of the transferred mass~($\beta$) is accreted and the remainder is lost from the system carrying away the specific angular momentum of the accretor \citep{Soberman+1997}. 
It can be shown that under these assumptions the orbit widens irrespective of the chosen mass transfer efficiency (value of $\beta$), as long as the mass of the accreting black hole becomes large relative to the donor mass ($M_{\rm d}$). For low $\beta$ this implies the orbit widens as soon as {$\Mbh \gtrsim 0.76$} times the mass of the donor (see Appendix~\ref{app: orbital evolution} for the derivation).
This condition is typically met for the progenitors of systems that can form PISN mass gap systems. This is a robust result that is also valid for other binaries that do not evolve into BHs \citep{Renzo+2019}.

We thus conclude that lowering the mass transfer efficiency under these assumptions does not increase the number of PISN mass gap mergers.

We can furthermore consider what happens when mass is lost with a higher specific angular momentum. For example, mass lost from the outer Lagrangian point may be ejected to form a circumbinary disk. 
This mode of mass loss leads to rapid shrinking of the orbit for almost all variations of mass transfer efficiency (see Appendix~\ref{app: orbital evolution} for details). Test simulations show that most BHs plunge into the companion's envelope, unless the mass transfer efficiency is highly fine-tuned. It is unclear what is the fate of such systems.

In conclusion, we do not expect that variations in the mass transfer efficiency can significantly increase the number of PISN gap mergers. 


\subsubsection{Can we accrete even more?}
\label{ss: accrete even more}

The most massive black hole involved in a BBH merger in our simulations has a mass of $\Mbhmax \approx 90\Msun$. Can $\Mbhmax$ be taken as a robust upper limit or are there uncertainties that allow us to increase $\Mbhmax$ further?


During stable mass transfer we already assume a mass transfer efficiency of 100\%. However, in our CE accretion model, the BHs typically accrete less than 20\% of the companions mass (see also Figure \ref{fig:remMassBH}).  While the assumptions in our second model variation are already extreme, it is worthwhile to consider what happens if BHs are allowed to accrete even more during the CE inspiral phase.  

\begin{figure}[ht]
\centering
  \includegraphics[trim=0.cm 0cm 0cm 0cm, clip, width=0.49\textwidth]{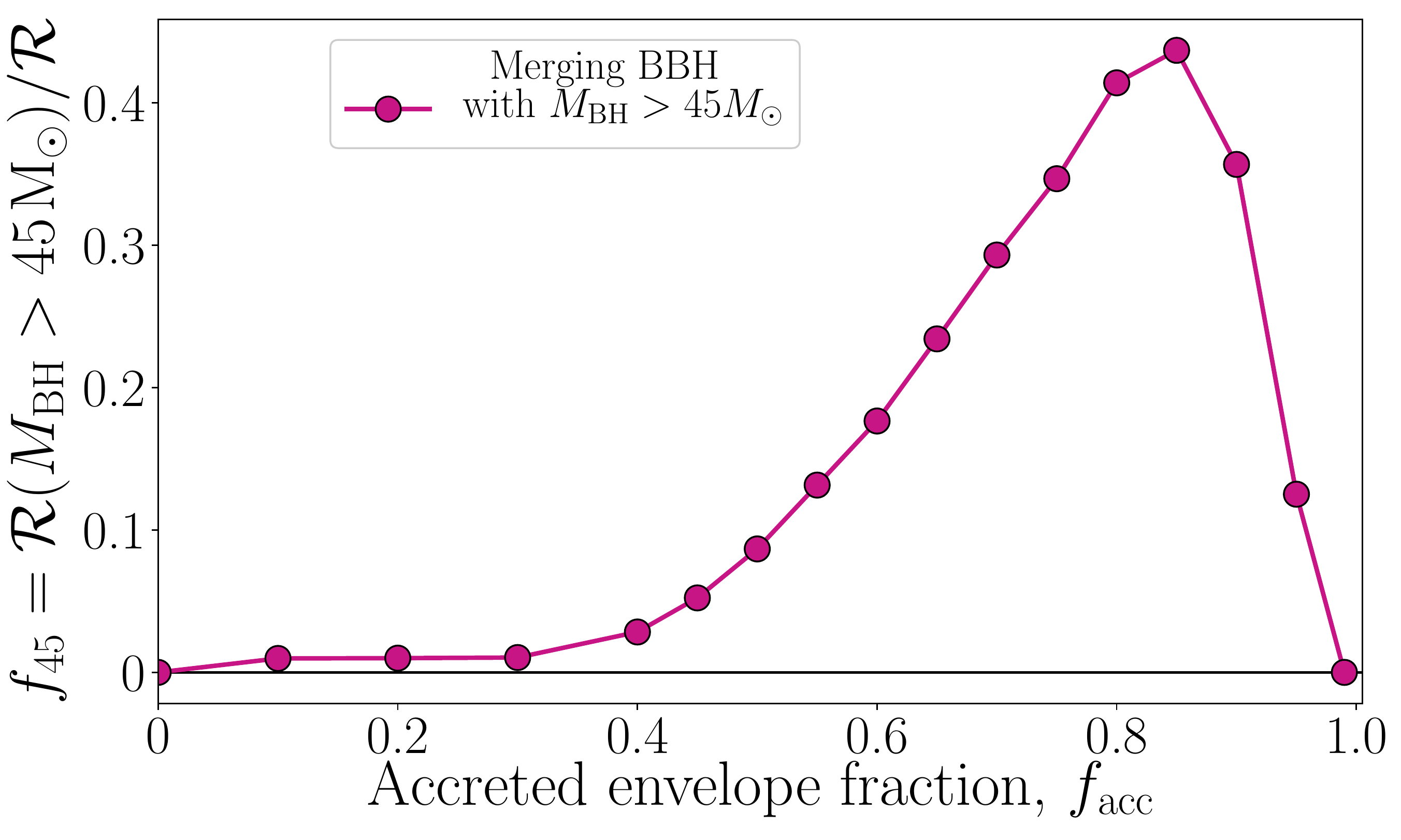} 
    \caption{The fraction of BBH mergers with a component mass $\Mbh > 45 \Msun$ (i.e. \fgalLIM), as a function of the fraction of the companion's envelope that is accreted during each CE mass transfer episode. Each point represents one simulation from the additional exploratory grid of 17 simulations discussed in Section \ref{ss: accrete even more}. } 
    \label{fig: optimalCEacc}
\end{figure}

To investigate this, we ran a grid of 17 additional simulations with the same setup as for our second model variation (as described in Section~\ref{sec:method}), but now assuming that a fixed fraction $f_{\rm acc}$ of the envelope is accreted onto the BH. We vary $f_{\rm acc}$ between 0.0 and 0.99.  We still estimate the final separations by considering the binding energy of the envelope, after subtracting the mass accreted by the BH. These additional simulations are run at a lower resolution of $10^5$ systems per simulation.  

Figure \ref{fig: optimalCEacc} shows the fractional rate of PISN gap mergers, $\rm f_{gal, 45}$ as a function of $f_{\rm acc}$. We see that the rate of PISN gap mergers increases with $f_{\rm acc}$ and peaks when BHs are assumed to accrete about $85\%$ of their companions envelope. This simulation predicts as many as $42\%$ of BBH mergers from PISN mass gap systems. For even higher values of $f_{\rm acc}$, we see that there is not enough of the envelope left to sufficiently shrink the orbit. We note that such high rates for PISN gap mergers are already contradicted by first and second LIGO and Virgo observing run \citep{Abbott2018bPop}.

These simulations further show that we can only obtain a significant fraction of PISN gap mergers ($> 2\%$) when BHs accrete at least $35\%$ of their companion's envelope mass during the envelope inspiral. This would suggest accretion rates that significantly surpass the Hoyle-Lyttleton accretion rate \citep{MacLeod2015} during every CE event. We consider it unlikely that such high rates are physical.

We thus consider it very unlikely that $\Mbhmax$ can be significantly increased beyond the values as quoted in Tables \ref{tab: masses} and \ref{tab: tot masses} for systems originating from isolated binary evolution.

\subsection{Effects of super-Eddington accretion on the lower mass gap}
\label{ss: lower mass gap}
\begin{figure}
    \centering
    \includegraphics[width=0.45\textwidth]{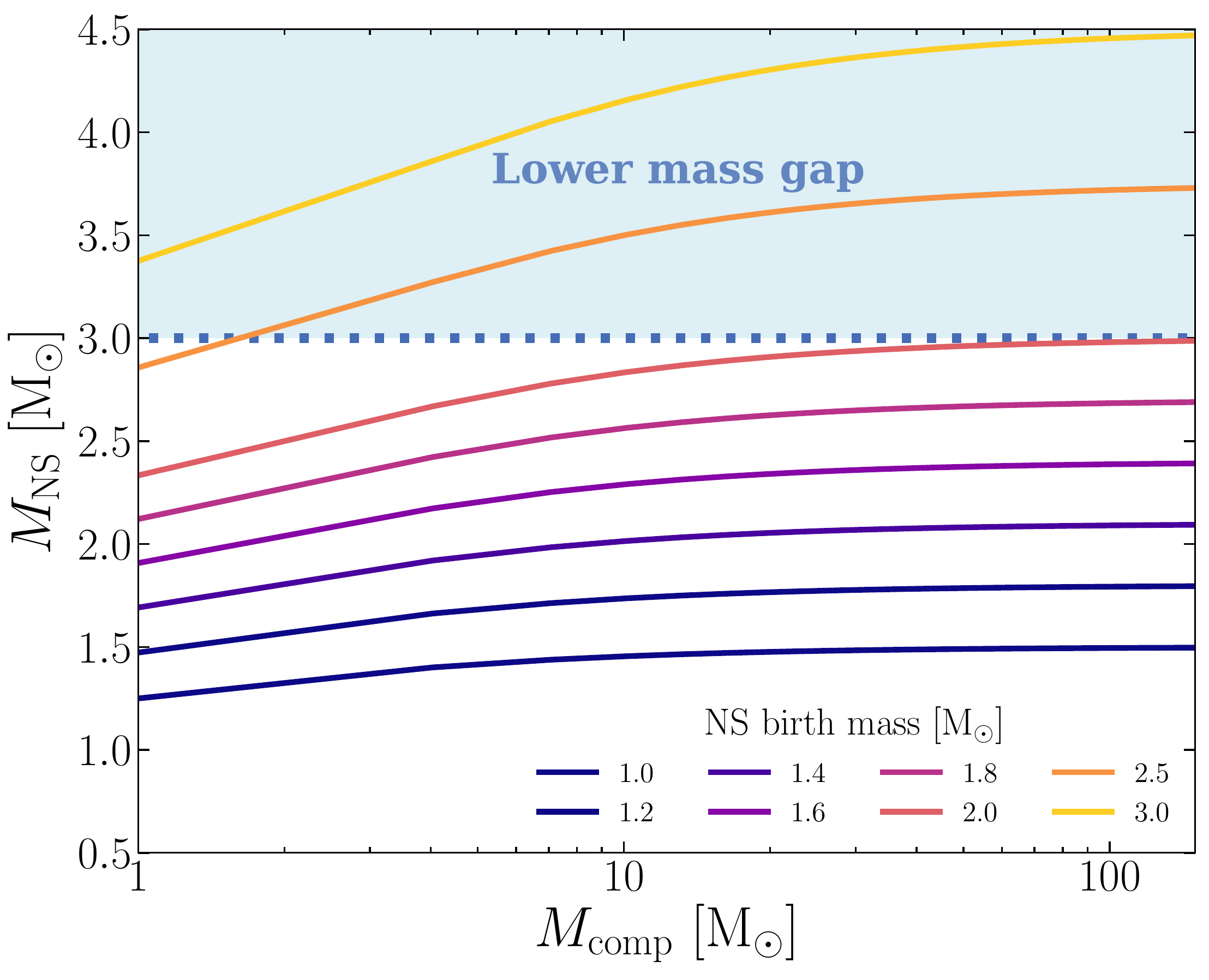}
    \caption{The maximum mass for an accreting NS according to Equation \ref{eq: BHL accretion rate} as a function of the companion mass at the moment of Roche-lobe overflow. This illustrates that only neutron stars that are born close to the lower mass gap will be able to accrete enough to end with a final mass in the lower mass gap. }
    \label{fig:remMassNS}
\end{figure}  

If all BHs that enter a CE-phase are allowed to accrete at a super-Eddington rate, we could hypothesize that the same accretion model would apply to neutron stars, as was originally suggested by \cite{Houck1991} \cite{Fryer+1996} and \cite{Popham1999}.
Allowing for accretion onto neutron stars during every CE event following Eq. \ref{eq: BHL accretion rate} could create an overabundance of BHs and possibly leads to BH neutron star and binary neutron star rates that are inconsistent with the current estimates from \cite{Abbott2018bPop}. 

We consider the consequences of super-Eddington accretion onto NS stars and whether this can populate the lower mass gap between neutron stars and BHs (between 3 and 5\Msun) by evaluating the effect of applying Equation \ref{eq: BHL accretion rate} to mass ranges that are relevant for neutron stars. 

In Figure \ref{fig:remMassNS} we show the maximum mass for accreting NS according to Equation \ref{eq: BHL accretion rate} as a function of the companion mass at the moment of Roche-lobe overflow. This illustrates that only neutron stars that are born close to the lower mass gap will be able to accrete enough to end with a final mass in the gap. 
We thus expect that extending our model variation 2 to mass ranges relevant for neutron stars, would not significantly pollute the lower mass gap, which is in accordance with the findings in \cite{MacLeod2015}.


\subsection{Caveats}
\label{ss: caveats} 

We emphasize that the simulations presented in this paper are extreme by design. We do not consider them realistic, but they are chosen to constrain the maximum amount by which the isolated binary evolutionary channel can pollute the PISN mass gap. 

Our results are subject to all caveats that apply to population synthesis simulations that make use of approximate evolutionary algorithms \citep[see e.g., ][]{Langer_2020}. Of primary concern is the treatment of the common-envelope phase, which is one of the least understood phases of binary interaction. 
\revision{A specific example is the stability of mass transfer in cases where the donor star evolves to become a convective red giant \citep[e.g.,][and references therein]{Pavlovskii_2015,Pavlovskii+2016}. }
The treatment we adopt in our second model variation is inspired by the results of hydro-dynamical simulations by \cite{MacLeod2015} and uses very simple scaling arguments as a recipe for accretion onto the compact object. The predictions for the final separation after the common envelope phase are based on simple energy arguments \citep{Webbink1984}. The treatment of this process is simplistic, and we hope this will be reconsidered carefully in future work as the understanding of the common-envelope phase increases. 

We are further affected by uncertainties in massive star evolution. The main open questions concern the role of stellar wind mass loss \citep[e.g.,][]{Smith2014, Renzo+2017}, and interior mixing \citep[e.g.,][]{Maeder+2000}, which affect the final core masses and radial evolution. The algorithms used in our simulations are based on detailed evolutionary simulations for single stars with masses up to 50\Msun. Above 50\Msun we rely on extrapolations of fits. 

\section{Conclusions and summary} 
\label{sec:conclusions}
In this work we investigate the BBH population in the pair-instability mass gap due to isolated binaries, by allowing for accretion onto BHs at a super-Eddington rate from their stellar companions.
We accomplish this by means of the population synthesis code \revision{\COMPAS}.

We place an upper limit on the contribution of isolated binaries to creating PISN mass gap events, defined as BBH mergers that contain a component with $\Mbh>~45\Msun$. 
We find that a substantial population of BBH systems with a component in the PISN mass gap can be formed via stable super-Eddington accretion onto BHs (see Figure \ref{fig:m1m2}). However, these systems will not contribute to the BBH merger rate since their binary orbits are typically too wide to merge within a Hubble time (Table \ref{tab: merger rates}).

In our most optimistic model, which allows for accretion onto BHs during both stable mass transfer and during a CE phase (model 3, combined), we find that less than about $2\%$ of all BBH mergers are expected to contain one component in the PISN mass gap (see Table \ref{tab: masses}). Moreover, only about $0.5\%$ of the merging BBH systems in this model variation have a total mass $\Mbhtot \geq 90\Msun$ (see Figure \ref{fig:masshist} and Table \ref{tab: tot masses}).
By design this model includes extreme assumptions about the accretion physics. More conventional assumptions significantly lower these fractions.\\

Our results show that the classical isolated binary formation scenario of BBHs is not expected to significantly pollute the pair-instability mass gap when compared to other pathways proposed in the literature (see Figure \ref{fig: compare to others}).
None of our simulations produce a merging BBH system with a total mass $\Mbhtot\geq~100\Msun$ (Table \ref{tab: tot masses}). 

\revision{We argue that BBH systems with $\Mbhtot \leq 100\Msun$ and extreme mass ratios, combined with  an aligned spin orientation could be indicative of BHs that underwent super-Eddington accretion from a companion star (Figure~\ref{fig:cartoon and table}).} However, at the time of writing, predictions from different pathways for the maximum masses, mass ratios and BH spins are not sufficiently constrained to decisively differentiate between pathways. 

We predict that the BBH population cannot be explained by the isolated binary evolution channel alone if more than $1\%$ of all BBH mergers has a mass higher than $90\Msun$. Future detections of PISN mass gap events will enable us to determine the relative contribution of different channels to the overall population of BBHs.\\

Our finding that the isolated binary evolutionary scenario does not introduce significant uncertainties for the existence and location of the PISN mass gap are promising. This strengthens the predictive power that can be drawn from \Mbhmax for constraining the relative contribution of different formation scenarios \citep{sedda2020fingerprints}, the physics of the progenitors including nuclear reaction rates  \citep{Farmer+2019}, and possibly even the Hubble constant \citep[e.g.][]{Farr2019}.


\acknowledgments {
We thank Y. G\"{o}tberg, M. MacLeod, E. Ramirez-Ruiz, R. Narayan, C. Rodriguez, N. Ivanova and the members of the \COMPAS collaboration for insightful discussions. This project was funded in part by the European Union's Horizon 2020 research and innovation program from the European Research Council (ERC, Grant agreement No. 715063), and by the Netherlands Organization for Scientific Research (NWO) as part of the Vidi research program BinWaves with project number 639.042.728. RF is supported by an NWO top grant (PI: de Mink) with project number 614.001.501.
}

\bibliography{my_bib, bib2}
\bibliographystyle{aasjournal}


\appendix

\section{\textbf{Angular momentum loss during (non) conservative mass transfer}}
\label{app: orbital evolution}
To evaluate the evolution of the binary separation during mass transfer, we need to quantify the angular momentum that is lost from the system. For this purpose we follow classical arguments describing the details of mass transfer in binaries,  (e.g. those presented in \citealt{van-den-Heuvel1994} and similarly Section 4 from \citealt{Renzo+2019}, and references therein).

The orbital evolution of binaries is well constrained by the change in total orbital angular momentum, J:

\begin{equation}
\label{eq: J}
\rm J^{2} = G \frac{M_d^{2} M_a^{2}}{M_d + M_a} a (1-e^{2})
\end{equation}
with $\rm G$ the gravitational constant, $\rm a$ the orbital separation, $\rm e$ the eccentricity and $\rm M_{d}$ the mass of the donor star, annd $\rm M_{BH}$ the mass of the accreting BH. 

We parametrise the amount of mass lost from the system with a conservativeness parameter $\beta$, defining $\rm \dot{M}_{BH} = -\beta \ \dot{M}_{d}$ where $\rm \dot{M}_{BH}$ and $\rm -\dot{M}_{d}$ are the mass accretion and donation rates respectively. We furthermore approximate the specific orbital angular momentum of the ejected matter as $\gamma$ times the specific angular momentum of the binary. The specific angular momentum of the ejected matter, $\rm h_{loss}$, can then be rewritten in terms of $\gamma$ and $\beta$:

\begin{equation}
\begin{split}
\label{eq: hloss}
\rm h_{loss} = \gamma \frac{J}{M_1 + M_2 } =  \frac{\dot{J}}{\dot{M}_1 + \dot{M}_2 } , \\
\rm \rightarrow \frac{\dot{J}}{J } =  \frac{\gamma (1 - \beta)\dot{M}_{d}}{M_d + M_a }.
\end{split}
\end{equation}

Using Eq. \ref{eq: J} we can derive a very general formula for the change in angular momentum:

\begin{equation} 
\rm 2 \frac{\dot{J}}{J} = 2 \frac{\dot{M}_d}{M_d} + 2\frac{\dot{M}_a}{M_a} -  \frac{\dot{M}_d + \dot{M}_a }{M_d + M_a} + \frac{\dot{a}}{a} + \frac{(-2 e \dot{e} )}{(1-e^2)}
\end{equation}

In the case of Roche-lobe overflow we assume the orbit is fully circularised, and thus the last term is zero \citep[see e.g.][for an expression of the orbital evolution incuding eccentricity]{Soberman+1997}.
Substituting the result from Eq. \ref{eq: hloss} and the definition of $\beta$, we can write this for the orbital evolution;

\begin{equation}
\label{eq: sep evol}
\rm \frac{\dot{a}}{a} =  -2 \frac{\dot{M}_{d}}{M_{d}} \left \{ 1 - \beta \frac{M_{d}}{M_{BH}} -(\gamma + \frac{1}{2})(1 - \beta) \frac{M_{d}}{M_{d} + M_{BH} }   \right \}
\end{equation}
or, rewriting to explicitly show the dependence on our different parametrisation parameters:

\begin{equation}
\label{eq: widen cond}
\rm \frac{\dot{a}}{a} =  -2 \frac{\dot{M}_{d}}{M_{d}} \left \{ 1 - f(\beta,M_{BH},M_{d}, \gamma)  \right \}
\end{equation}

Since $\rm \dot{M_d} < 0$, we see that the orbit shrinks ($\rm \dot{a} < 0$) as soon as $\rm f(\beta,q, \gamma)$ is larger than one.

\begin{figure}[ht]
    \centering
    \includegraphics[width=0.5\textwidth]{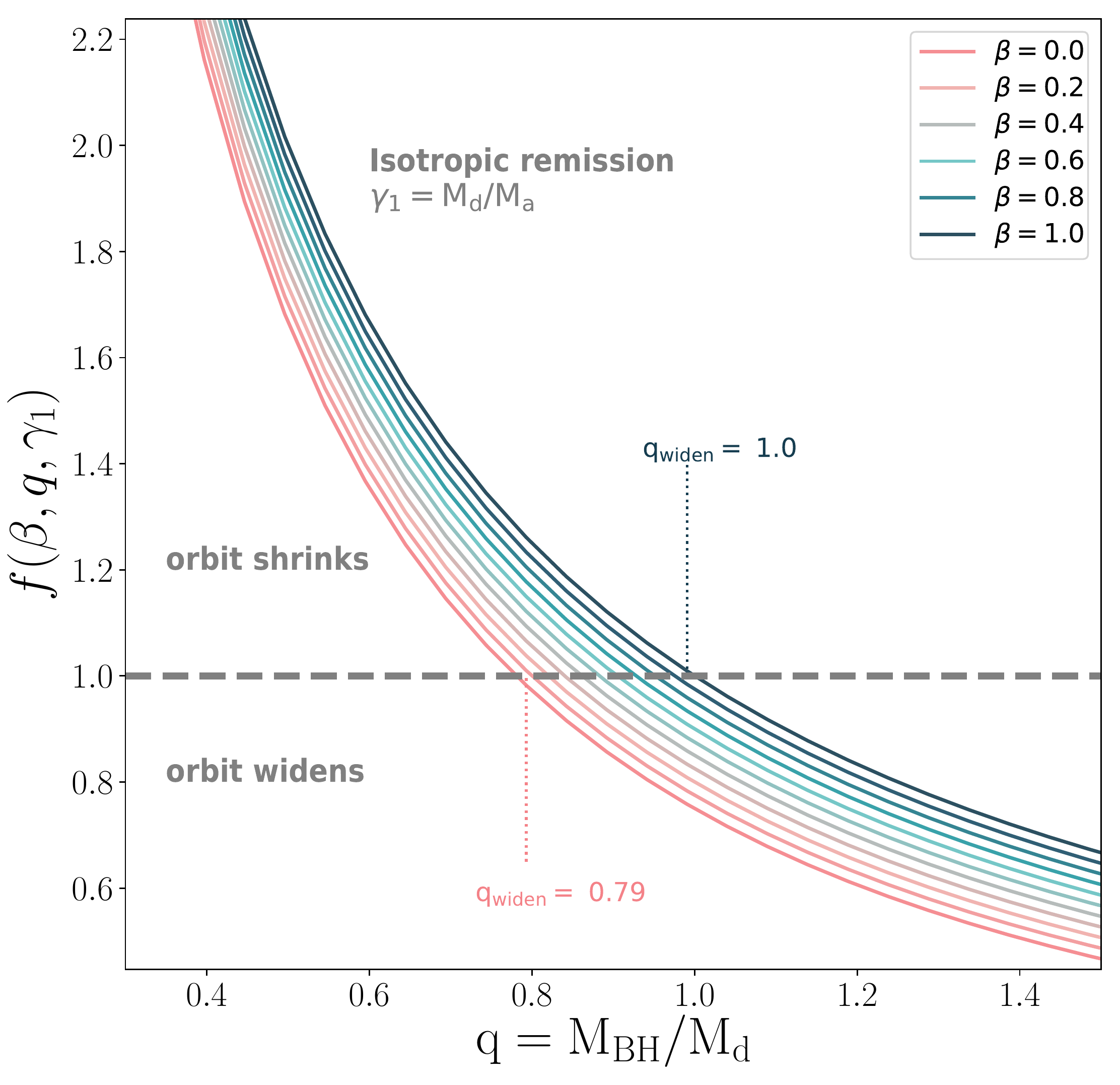}
    \caption{Parametrisation of the specific orbital angular momentum evolution that determines whether mass transfer will shrink (above the dashed black line) or widen (below the dashed black line) the binary orbit. The conservativeness parameter $\beta$ varies from completely non-conservative ($\beta=0$) to completely conservative ($\beta=1$). This assumes isotropic reemission of the ejected matter, i.e., assuming the ejected matter carries the specific angular momentum of the accretor. } 
    \label{fig: orbitEv}
\end{figure}   

It is now a matter of specifying $\gamma$ and $\beta$. 

\subsection{Varying the mass conservation}
In this work we assume isotropic reemission of the ejected matter. This assumes that the mass is ejected from the vicinity of the accretor, e.g. when mass is lost via bipolar outflows from a compact object. In this case we can approximate $\gamma_1 \rm  = M_{d}/M_{BH}$ \citep[e.g.][]{Soberman+1997}.
We can rewrite Equation \ref{eq: widen cond} as

\begin{equation}
\label{eq: wid isotrop}
f(\beta, q, \gamma_1) = \frac{\beta}{q} + \left( \frac{1}{q} + \frac{1}{2} \right) \left(1 - \beta \right) \frac{1}{q + 1}
    \begin{cases}
    > 1 & \textbf{orbit shrinks}, \ (\dot{a} < 0)  \\
    < 1 & \textbf{orbit widens},  \ (\dot{a} > 0)
    \end{cases}
    ,
\end{equation} 
where we have used $\rm q \equiv M_{BH}/M_{d}$ to describe the mass ratio.

Figure \ref{fig: orbitEv} displays the condition for widening or shrinking the orbit (Eq. \ref{eq: wid isotrop}) as a function of the mass ratio and the conservativeness parameter.
In the case of stable mass transfer from a star onto a BH, the binary \revision{typically} starts with $q \leq 1$ and then moves to higher $\rm q$. For fully non-conservative mass transfer ($\beta = 0.0$) the system will widen as soon as the mass ratio $\rm q > q_{widen} = 0.76$. For fully conservative mass transfer ($\beta = 1.0$) the system will widen as soon as the mass ratio $\rm q > q_{widen} = 1.0$. \revision{Systems where the accreting BH is more massive than the donor at the start of the stable mass transfer will always widen.} \\

Figure \ref{fig: orbitEv} shows that mass transfer \textit{will always widen the orbit} when the mass ratio $\rm q$ is larger than some mass ratio $q_{\rm widen}$, as long as the lost mass is presumed to carry the specific angular momentum of the accretor.
We therefore expect high amounts of mass transfer to always lead to significant widening of the binary system, regardless of the conservativeness of the mass transfer, and thus prevent the BBH from merging within a Hubble time.\\

In model 1, allowing for stable super Eddington accretion onto BHs, we assume the mass transfer is completely conservative ($\beta = 1$). Based on Figure \ref{fig: orbitEv} we conclude that varying the mass conservation in model 1 would still lead to significant widening of the BBHs and thus would not change our main conclusions.

\subsection{Varying the specific angular momentum lost}
If the mass that is lost from the system carries sufficiently high specific angular momentum, the orbit will shrink.  
If the lost mass ends up in a Keplerian orbit around the binary, it is called a circumbinary ring. This may occur when mass escapes through the outer Lagrangian point, L2.
The angular momentum of such a ring would correspond to \citep{Artymowicz1994}:
\begin{equation}
\rm \gamma_2 = \frac{(M_{d} + M_{BH})^{2}}{M_{d}M_{BH}} \sqrt{\alpha},
\end{equation}{}
where $\alpha = \rm a_{ring}/a$ is the ratio of the orbital separation of the binary over the distance between the circumbinary ring and the center of mass. For typical parameter of viscous disks, the location of the inner edge of a circumbinary disk varies between $1.8a$ and $2.6a$ \citep{Artymowicz+1994}.

Equation \ref{eq: widen cond} can then be rewritten as:
\begin{equation}
f(\beta, q,\gamma_2) = \frac{\beta}{q} + \frac{(1+q)(1 - \beta)}{q}\sqrt{\alpha} + \frac{1-\beta}{2(q + 1)} \sqrt{\alpha}
    \begin{cases}
    > 1 & \textbf{orbit shrinks}, \ (\dot{a} < 0)  \\
    < 1 & \textbf{orbit widens},  \ (\dot{a} > 0)
    \end{cases}
    .
\end{equation} 

We have calculated the effect of this mode of non-conservative mass transfer for $\alpha = 2$. Larger $\alpha$ lead to higher $\rm \dot{a}$, though varying between $1.8 a$ and $2.6 a$ has little effect. 
We find that non-conservative mass transfer ($\beta \leq 0.3$) in combination with high mass-transfer rates leads to shrinking of the binary orbit in a runaway fashion, which leads to a stellar merger. The orbit will shrink with increasing speeds (increasing $f(\beta, q,\gamma_2)$) as the mass ratio q increases.

For mass transfer with slightly to highly conservative mass transfer ($\beta \geq 0.3$) it is unclear what the fate of the systems will be. Test simulations using $\alpha = 2$, indicate that most BHs in this situation will plunge into their companion’s envelope,  unless the mass transfer efficiency is highly fine-tuned. More detailed simulations of this specific scenario are needed to determine its plausibility.

\section{\textbf{BH Formation yields and merger rates}}
\label{app: merger rate}
Our calculation of the BH formation yields follow \cite{Dominik+2012} and \cite{Neijssel+2019} but includes the weights from the adaptive sampling \citep[as described in][]{Broekgaarden2019}.

We start by calculating the total stellar mass contained in a synthetic galaxy ($\rm M_{\star, gal}$), assuming a \cite{Kroupa2001} initial mass function with initial masses in the range $0.08 - 200 \Msun$. We compute the subset of this synthetic galaxy that is spanned by our set of initial parameters ($\rm M_{\star, sub-gal}$) by integrating over the volume of initial parameter space. In our simulations, we adopt a binary fraction $f_{\rm bin} = 0.5$ \citep[e.g. ][]{Sana+2013} and draw initial masses in the range $20 - 150 \Msun$. The fraction of the synthetic Universe that is spanned by our initial parameter space is now computed as follows:  

\begin{equation}
    \rm f_{sim} = \frac{ M_{\star, sub-gal}}{ M_{\star, gal}}.
\end{equation}

The total star forming mass that our simulation represents is given by:
\begin{equation}
    \rm M_{SF} = M_{sim} \cdot f_{sim}^{-1} ,
\end{equation}
where $\rm M_{sim}$ is the total initial mass that is evolved in \COMPAS. 
This is used to calculate the number of merging BBHs formed per unit of star forming mass:

\begin{equation}
\label{eq N_Msf}
    \rm \frac{N_{BBH, t_H}}{M_{SFR}} = \frac{\sum^{\revision{N} }_{i} \delta_{t_{H},i} w_i }{ M_{SF}}, \hspace{1cm} with \hspace{0.1cm} \delta_{t_{H},i} = 
\begin{cases}
 1, & \revision{\text{if } type(i)\text{= BBH, and }\rm t_{delay,i} < t_{H}  }\\
 0,& \rm  \revision{otherwise}
\end{cases}
.
\end{equation}
\revision{Here $\rm N_{BBH, t_H}$ is the total number of BBHs formed with a coalescence time that is less than the Hubble time, $\rm \delta_{t_{H},i}$ is the Dirac delta function that equals 1 for a BBH system with a coalescence time that is less than the Hubble time, i.e. if it merges within the age of the Universe, and N is the total number of samples in the simulation N$ = 1\times 10^{6}$.} Finally, $\rm w_i$ is the formation weight of the binary based on the adaptive importance sampling algorithm as described in \cite{Broekgaarden2019}.
We estimate the \revision{absolute} 1--$\sigma$ statistical sampling uncertainty on the number of BBHs that merge in a Hubble time $\rm N_{BBH, t_H}$ \revision{ by computing the variance about the mean, that is
\begin{equation}
\label{eq: sigmaN}
\rm \sigma^2 \approx   \sum^{N}_{i}(\delta_{t_{H},i}^2 w^2_i) - \frac{\left[\sum^{N}_{i}(\delta_{t_{H},i} w_i)\right]^2 }{N} , 
\end{equation} 

}

We calculate merger rates for a synthetic galaxy following the same procedure as \cite{Belczynski+2016} and \cite{de-Mink+2015}. For this purpose, we calculate the number of coalescing BBHs occurring in a synthetic galaxy, observed per Myr today;
\begin{equation}
    \rm N_{BBH,gal} = \frac{N_{BBH,10}}{M_{SFR}} \cdot SFR_{gal} \cdot t_{gal},
\end{equation}
with a constant star formation rate, SFR$_{\rm gal} = 3.5 \Msun$ yr$^{-1}$ , and a galaxy lifetime $\rm t_{gal} = 10$ Gyr. These properties are chosen to resemble the Milky Way \citep[following estimates from][]{Flynn+2006,McMillan2011}.
The number of merging BBH systems per unit star forming mass, $\rm N_{BBH,10}/M_{SFR}$, is defined in a similar way as in Equation \ref{eq N_Msf}, but we now require the BBHs to merge in less than the age of the galaxy, $t_{\rm gal} = 10\Gyr$. \revision{The statistical sampling uncertainty on $N_{BBH,10}$ is estimated analogous to equation \ref{eq: sigmaN}}.

The merger rate per synthetic galaxy is then calculated as
\begin{equation}
    \Rmwg  \rm = \frac{N_{BBH,gal}}{t_{gal} } \ [Myr^{-1}],
\end{equation}
note that the age of the synthetic galaxy $t_{\rm gal}$, cancels out in this equation. $t_{\rm gal}$ only appears in the equivalent of Equation \ref{eq N_Msf}, when calculating the number of merging BBH systems per unit star forming mass. 

The merger rate per synthetic galaxy can be converted into an approximate volumetric rate following:
\begin{equation}
\Rvol \rm = 10^3 \left[ \frac{\rho_{gal} }{\rm Mpc^{-3}} \right] \left[ \frac{\Rmwg  }{\rm Myr^{-1}} \right]  \ yr^{-1} Gpc^{-3}
\end{equation}
where $\rm \rho_{gal} = 0.0116$ Mpc$^{-3}$ is the local density of Milky Way-like galaxies \citep[e.g.][]{Kopparapu2008}.

All uncertainty ranges on the formation yields and merger rates as quoted in this work are estimates of the 1--$\sigma$ statistical sampling uncertainty following from Equation \ref{eq: sigmaN}.



\section{\textbf{Remnant mass prescription}}
\label{app: farm fry}


To calculate the remnant masses we adopt the delayed model from \cite{Fryer+2012} for estimated CO core masses at the moment of core collapse $\Mco < 30 \Msun$, while we follow \cite{Farmer+2019} for $\Mco > 30 \Msun$.

Previous works studying the PISN gap \citep{Stevenson+2017, Belczynski+2016} use \cite{Fryer+2012} for $\Mco > 30 \Msun$.
\cite{Fryer+2012} compute the remnant mass based on the estimated helium core masses at the moment of core collapse while \cite{Farmer+2019} account for a PISN and compute the remnant mass based on the estimated CO core masses at the moment of core collapse.
Mapping between the helium core masses and PISN depends on uncertain physics such as the efficiency and extent of mixing (overshooting) which varies between models and with wind mass loss. 
The CO core mass at the moment of supernova is therefore a more robust parameter than the helium core mass at the moment of supernova to map the pre-supernova stellar properties to the final remnant mass \citep{Farmer+2019}. 

For $\Mco > 30 \Msun$ the prescription from \cite{Farmer+2019} results in lower remnant masses with respect to the prescriptions from \cite{Fryer+2012}. This is because \cite{Fryer+2012} does not account for pulsational pair-instability supernovae. The maximum BH mass formed in our simulation at $Z = 0.001$ for \cite{Farmer+2019} is $\Mbhmax = 43.4 \Msun$, while $\Mbhmax = 54 \Msun$ for \cite{Fryer+2012}.

\section{\textbf{Additional material}}
\label{app: additional material}

For each model variation, we provide a \python file describing the initial conditions as used in each of our 4 model variations described in Section \ref{sec:method}, and a HDF file containing i.a. a list of compact object properties as resulting from our \COMPAS simulations at \url{https://doi.org/10.5281/zenodo.3746936} and \url{https://liekevanson.github.io/IsolatedBinaries_PISNgap.html}.

\end{document}